\documentclass[aps,prd,reprint,longbibliography,nofootinbib,superscriptaddress]{revtex4-2}

\usepackage{natbib}
\usepackage{slashed}
\usepackage{verbatim}
\usepackage[T1]{fontenc}
\usepackage{mathbbol}
\usepackage[dvipsnames]{xcolor}
\usepackage{orcidlink}
\usepackage[english]{babel}
\usepackage{lipsum}
\usepackage{physics}
\usepackage{array}
\usepackage{dcolumn}
\usepackage[normalem]{ulem}
\usepackage{url}
\usepackage{tensor}
\usepackage{comment}
\usepackage{graphicx,color,overpic,mathtools}
\usepackage{amsthm,amsmath,amssymb,mathrsfs}
\usepackage{soul}
\usepackage{microtype}
\usepackage{braket,bm,bbm,setspace}
\usepackage{cancel}
\usepackage{float}
\usepackage{xargs}
\definecolor{teal}{RGB}{0, 128, 128}
\definecolor{myred}{RGB}{179, 27, 27}
\usepackage{hyperref}
\hypersetup{
    colorlinks=true,
    linkcolor=myred, 
    citecolor=myred, 
    urlcolor=myred  
 }

\def\be{\begin{equation}}
\def\ee{\end{equation}}

\newcommand{\beq}{\begin{eqnarray}}

\newcommand{\eeq}{\end{eqnarray}}

\def\be{\begin{equation}}
\def\ee{\end{equation}}
\def\bea{\begin{eqnarray}}
\def\eea{\end{eqnarray}}
\def\beq{\begin{eqnarray}}
\def\eeq{\end{eqnarray}}

\begin{document}
\title{Tidally-enhanced resonances in  extreme-mass-ratio inspirals:\\ A tertiary path to chaos}
\author{Kyriakos Destounis}
\affiliation{CENTRA, Departamento de Física, Instituto Superior Técnico – IST, Universidade de Lisboa – UL, Avenida Rovisco Pais 1, 1049-001 Lisboa, Portugal}
\affiliation{Theoretical Astrophysics, Institute for Astronomy and Astrophysics, University of T\"{u}bingen, 72076 T\"{u}bingen, Germany}
\affiliation{School of Applied Mathematical and Physical Sciences – SEMFE, Physics Department, National Technical University of Athens -- Polytechnic Campus, Zografou 15772, Athens, Greece}
\author{Takuya Katagiri}
\affiliation{Dipartimento di Fisica, Sapienza Università di Roma, Piazzale Aldo Moro 5, 00185, Roma, Italy}
\affiliation{INFN, Sezione di Roma, Piazzale Aldo Moro 2, 00185, Roma, Italy}
\author{Sajal Mukherjee}
\affiliation{Birla Institute of Technology and Science Pilani, Rajasthan, 333031, India}
\author{Kostas D. Kokkotas}
\affiliation{Theoretical Astrophysics, Institute for Astronomy and Astrophysics, University of T\"{u}bingen, 72076 T\"{u}bingen, Germany}

\begin{abstract}
Extreme-mass-ratio inspirals (EMRIs) provide a unique laboratory for probing strong-field gravity and complex relativistic dynamics. We study a tidally deformed EMRI composed of a stellar-mass secondary orbiting a supermassive (non-)rotating black hole embedded in an external, adiabatically varying tidal environment. These systems provide a restricted, yet astrophysically motivated, realization of the relativistic three-body problem, expected to appear in active galactic nuclei, with a clear hierarchy of masses and radiation-reaction timescales. The external tidal deformation breaks the axisymmetry of the Kerr spacetime, rendering the geodesic dynamics non-integrable and giving rise to chaotic motion. The resulting signature of non-integrability is characterized through Poincar\'e maps and rotation curves constructed from the ratios of the fundamental frequencies of bound radial, polar, and azimuthal motion. We identify two prominent plateaus whose widths increase with the tidal field amplitude, signaling a transition from weak to strong chaos. We then demonstrate the sensitivity of the chaotic dynamics to the orientation of the orbit relative to the tidal field. We further analyze the proper-time evolution of the action-angle variables, showing that the angle combinations associated with the dominant commensurabilities are phase locked, thereby allowing the associated tidal contributions to induce secular changes in the constants of motion, whereas off-plateau angle combinations circulate. These results clarify the dynamical significance of the prominent plateaus and provide a novel phase-space characterization of tidal resonances in EMRIs. Finally, we discuss the potential role of radiation-reaction effects in driving EMRIs through tidal island crossings and the implications for gravitational-wave inference with future detectors.
\end{abstract}

\maketitle

\tableofcontents
\section{Introduction}

We currently live in the era of gravitational-wave (GW) astronomy \cite{Bailes:2021tot} and multimessenger astrophysics \cite{Meszaros:2019xej}. GW detections are streamlined and continuously upgraded to achieve higher precision and accuracy in the kiloHertz (kHz) frequency domain that ground-based interferometers, such as the LIGO/Virgo/KAGRA (LVK) collaboration, operate \cite{KAGRA:2021vkt}. The current LVK-detected signals are directly intertwined with the physical properties of the source components before the collision of binary black holes (BHs). In turn, the final GW vibrations of the BH remnant is currently studied with the modal tools of BH spectroscopy and other complementary non-modal notions \cite{Kokkotas:1999bd,Dreyer:2003bv,Berti:2009kk,Cardoso:2019rvt,Hughes:2019zmt,Destounis:2023ruj,Berti:2025hly,Jaramillo:2020tuu,Jaramillo:2022kuv,Destounis:2021lum,Destounis:2023nmb,Sarkar:2023rhp,Boyanov:2022ark,Boyanov:2023qqf,Cheung:2021bol,Boyanov:2022ark,Boyanov:2024fgc,Cai:2025irl,Rosato:2024arw,Ianniccari:2024ysv,MalatoCorrea:2025iuc,Destounis:2025dck,dePaula:2025fqt,Cheung:2022rbm,Mitman:2022qdl,Sberna:2021eui,Kehagias:2023ctr,Perrone:2023jzq,Kehagias:2024sgh}. These precise measurements define the pinnacle of GW inference and help us trace the darkest corners of the Cosmos.

The triumphant success of the LVK collaboration has set new frontiers in GW astronomy \cite{Berti:2015itd}. Probing the kHz domain of GWs is only the beginning of what will become a long and remarkable era, in which we are currently setting the scene in order to access the milliHertz (mHz) domain of GWs. The spaceborne detector \emph{Laser Interferometer Space Antenna (LISA)} \cite{LISA:2017pwj} has been adopted and funded by the European Space Agency, together with NASA, and collectively with the proposals of TianQin \cite{Li:2024rnk} and Taiji \cite{Ruan:2018tsw,Ren:2023yec} from China, it will grant us access to sources that span from local galactic binaries to global cosmological-scale sources of mHz GW radiation \cite{Barausse:2020rsu,LISA:2022yao,LISA:2022kgy,Karnesis:2022vdp}.

One of the most interesting targets of space-based detectors is highly-asymmetric binaries, known as extreme-mass-ratio inspirals (EMRIs). These binaries, which consist of a supermassive dark primary and a stellar-mass compact object secondary, have such a huge mass disparity ($10^4$ to $10^7$ orders of magnitude) that they will last for years, perform $10^4-10^5$ revolutions in the strong-field regime, and will produce GWs that peak at the mHz frequency domain of the LISA detector. Eventually, their observation will enable even more precise GW inference and will unveil new phenomena \cite{Amaro-Seoane:2007osp,Babak:2017tow,Berry:2019wgg,Cardenas-Avendano:2024mqp}, that have not been observed yet due to the bandwidth limitations of ground-based detectors \cite{Barack:2003fp,Baibhav:2019rsa}.

Observations of deep space and the Universe as a whole reveal an overall Erebos-shrouded expanse mostly dominated by vacuum. Nevertheless, the standard model of Cosmology points to an expanding Universe filled with faint radiation, interstellar dust, magnetic fields, and dense astrophysical environments, such as accreting baryonic matter around BHs and dark matter halos of all scales \cite{Bovy:2015mda,Scognamiglio:2026phv,Cardoso:2025npr}. Accretion disks around BHs are extremely dense, especially in stellar-mass BHs where high-density plasma forms \cite{Deprince:2020fxe,Jiang:2019xqn,Jiang:2019ztr}. Galactic centers host massive, luminous, and turbulent structures of plasma that, together with galactic supermassive BHs, power active galactic nuclei \cite{Szuszkiewicz:2001} which eventually form EMRIs surrounded by accretion disks \cite{Kocsis:2011dr,Duque:2024mfw,HegadeKR:2025dur,Zwick:2025wkt,Duque:2025yfm,Spieksma:2025wex}. Besides known baryonic matter environments \cite{Richards:2015gla}, dark matter halos are theorized to enshroud galaxies \cite{Casetano:1983,Karukes:2015fma,Fraternali:2011vm}, thus solving the galactic rotation curve problem \cite{Jimenez:2002vy,Villano:2022cry}, and even stabilizing clusters of galaxies with very massive dark matter coronae \cite{Einasto:1974dra}. Intriguingly, even though the cosmic dark matter density is considered low, galaxies and galactic clusters host densities that are considerably higher \cite{Bertone:2024rxe}. Therefore, EMRIs in galactic dark matter have to be studied with relativistic methods at the spacetime \cite{Boehmer:2007az,Cardoso:2021wlq,Konoplya:2021ube,Konoplya:2022hbl,Cardoso:2022whc,Destounis:2022obl,Speeney:2022ryg,Rahman:2023sof,Dai:2023cft,Jusufi:2022jxu,Figueiredo:2023gas,Speeney:2024mas,Datta:2023zmd,Pezzella:2024tkf,Ovgun:2025bol,Gliorio:2025cbh,Myung:2024tkz,Zhang:2024hjr,Xavier:2023exm,Macedo:2024qky,Maeda:2024tsg,Shen:2023erj,Shen:2024qbb,Kouniatalis:2025itj,Datta:2025ruh,Datta:2026krm,Fernandes:2025lon,Fernandes:2025osu,Destounis:2025tjn} and the dynamics level \cite{Sadeghian:2013laa,Boskovic:2018rub,Vicente:2019ilr,Annulli:2020lyc,Kavanagh:2020cfn,Coogan:2021uqv,Vicente:2022ivh,Traykova:2023qyv,Duque:2023seg,Karydas:2024fcn,Kavanagh:2024lgq,Mitra:2025tag,Vicente:2025gsg} in order to appropriately capture environmental effects in BH binaries \cite{Barausse:2014tra,Cardoso:2019rou,Toubiana:2020drf,Zwick:2021dlg,Zwick:2022dih,Sberna:2022qbn,Cole:2022yzw,CanevaSantoro:2023aol,Boskovic:2024fga,Blas:2024duy,Roy:2024rhe,Dyson:2025dlj}.

Despite the fact that non-vacuum BH binaries are now in the spotlight, due to the Universe veiling them with matter, the paradigm that the formation of BH binaries requires tertiary bodies has been physically motivating, and recently observed \cite{Mazzolari:2026A&A}, thus leading to three-body dynamics. This concept falls under the form of binary EMRIs or relativistic triple systems of similar-mass components \cite{Dmitrasinovic:2014lha,Meiron:2016ipr,Stegmann:2025zkb,Dittmann:2023sha,Yang:2024tje,Santos:2025ass,Cardoso:2026ugm,Klimesova:2026rme}. The potential presence of additional bodies in the vicinity of compact binaries introduces environmental effects that complicate the construction of precise GW templates. Addressing this challenge requires, as a first step, a systematic exploration of the tidal dynamics experienced by gravitating bodies in strong external fields. Although analogous effects are well understood in the Newtonian limit—and have even enabled indirect constraints on unseen BHs in the center of the Milky Way \cite{Naoz:2019sjx,Will:2023nlt,GRAVITY:2023}—their manifestation in the relativistic regime remains comparatively unexplored. Here, we focus on tidal phenomena in strong-field gravity, extending previous analyses restricted to (non-)rotating BHs \cite{Cardoso:2021qqu,Camilloni:2023rra,Poisson:2014gka,Landry:2015zfa,Pani:2015hfa,Giataganas:2026gkl}. We further leverage this setting to investigate the structure of test-particle motion around tidally-deformed, rapidly-rotating BHs \cite{Thorne:1980ru,Leung:1997was,Binnington:2009bb,Cardoso:2020hca,Kyriazis:2025fis,Cocco:2026lkr,Katagiri:2026gkz}, and to assess how nearby tertiary companions can introduce chaotic phenomena in EMRIs; an aspect of restricted triple system tidal interactions that have been merely studied \cite{Barausse:2007dy,Tokovinin:2006jm,Antonini:2012ad,Antonini:2015zsa,Meiron:2016ipr,Stone:2016wzz,Bartos:2016dgn,Robson:2018svj,Han:2018hby,Chen:2018axp,Gupta:2019unn,Martinez:2020lzt,Gupta:2021cno,Liu:2020gif,Cardoso:2021vjq}.

In this work, we study the phase space of geodesic motion around a (non-)rotating BH in the tidal field of a distant third body 
as a model for EMRIs in a tidal environment.  The present work extends several representative results of the phase-space analysis of Ref.~\cite{Katagiri:2026gkz}. Specifically, by specializing the scope to EMRIs, we analyze the phase-space structure over a broad range of system parameters. We find evidence of non-integrability of the geodesics and chaotic phenomenology, triggered by the third body's tide, in terms of Poincar{\'e} map and rotation curves. Such effect is known to create phenomenologically discernible effects to GWs from EMRI sources \cite{Destounis:2021mqv,Destounis:2021rko}. Since the current literature does not take into account the possibility of resonance enhancement in tidal EMRIs \cite{Bonga:2019ycj,Gupta:2021cno,Gupta:2022fbe}, here we address it in depth at the geodesic level. We then analyze the proper-time evolution of action-angle variables~\cite{Schmidt:2002qk} and show that the angle combinations associated with the dominant commensurabilities are phase locked, thereby allowing the associated tidal contributions to induce secular changes in the constants of motion.

In what follows, we use geometrized units such as $G=c=1$, unless stated otherwise.

\section{Tidally deformed rotating black holes} \label{sec:DeformedMetric}

\subsection{The metric}

In this section, we briefly introduce the tidally-deformed rotating BH metric. The deformed geometry is constructed perturbatively in terms of the tidal-field strength~\cite{Katagiri:2026gkz}. The zeroth-order spacetime is described by the Kerr geometry. In Boyer-Lindquist coordinates $(t,r,\theta,\phi)$, the Kerr metric~$g_{\mu\nu}^{\rm Kerr}$ reads
\begin{align}
    &g_{\mu\nu}^{\rm Kerr}dx^\mu dx^\nu=-\left(1-\frac{2Mr}{\Sigma}\right)dt^2-\frac{4M a r \sin^2\theta}{\Sigma}dtd\phi\\
    &+\frac{\Sigma}{\Delta}dr^2+\Sigma d\theta^2+\left(r^2+a^2+\frac{2M a^2 r\sin^2\theta}{\Sigma}\right)\sin^2\theta d\phi^2,\nonumber\\
  &\Sigma:=r^2+a^2\cos^2\theta,\quad \Delta:=r^2-2M r+a^2,
\end{align}
which is completely characterized by two parameters, mass~$M$ and spin~$a$. There exist two horizons at $r_\pm:=M\pm\sqrt{M^2-a^2}$. Henceforth, $r_+>r_-$ and $M>a$ should hold, where $r=r_+$ is the event horizon and $r=r_-$ is the boundary of maximal, globally-hyperbolic development of Cauchy initial data, i.e., the Cauchy horizon.

We place an external gravitational field around a Kerr BH. Assuming that the tidal field is sufficiently weak and varies in time adiabatically, the tidal interaction with the BH can be modeled within linear, stationary perturbation theory. The presence of the external tidal field induces a linear stationary response of the geometry, leading to tidal deformations of the BH. The resulting metric takes the schematic form,
\begin{align}
    g_{\mu\nu}=g_{\mu\nu}^{\rm Kerr}+h_{\mu\nu}.\label{eq:metric}
\end{align}
The second term $h_{\mu\nu}$ denotes the metric perturbation associated with the tidal deformation. The perturbed metric~$h_{\mu\nu}$ is constructed with the method outlined in Ref.~\cite{Katagiri:2026gkz}; examples with a variety of BH spin values $a/M$ are also provided online~\cite{Notebook,CoG}.

It is worth noting that $h_{\mu\nu}$ depends on $r$ and $\theta$, as well as the azimuthal angle. Therefore, the deformed geometry admits a time-translational Killing vector field, within the current framework, while the axisymmetry of Kerr spacetime is not preserved. The perturbed spacetime no longer belongs to the Petrov-type D class, indicating that hidden symmetries of the Kerr metric may be lost. The components of $h_{\mu\nu}$ grow with $r$; however, the magnitude of the perturbation remains bounded due to the hierarchy between the length scales of the tidal environment and the BH \cite{Katagiri:2026gkz}.

The metric for tidally-perturbed rotating BHs, constructed in Ref.~\cite{Katagiri:2026gkz}, is expressed in the advanced Kerr coordinates. In this work, we will adopt a simpler representation of the metric through the Boyer-Lindquist coordinates. To this end, we perform a coordinate transformation from the advanced Kerr coordinates~$(v,r_{\rm aK},\theta_{\rm aK},\varphi)$ to the Boyer-Lindquist coordinates $(t,r,\theta,\phi)$ by introducing
\begin{equation}
\begin{aligned}
    dv= dt+\frac{r^2+a^2}{\Delta} dr,\,dr_{\rm aK}=dr, \\
    d\theta_{\rm aK}=d\theta,\quad  d\varphi=d\phi+\frac{a}{\Delta}dr.
\end{aligned}
\label{eq:FromaKtoBL}
\end{equation}
The coordinate-frame notation \eqref{eq:FromaKtoBL} will be fundamental for the stable evolution of geodesics in tidally-deformed spacetimes.

\subsection{Characterizing external tidal fields}

Here, we specify the gravitational field that creates a tidal field leading to the deformation of the Kerr geometry. For EMRIs, we assume that the deformed massive BH is a component of a binary system in a (quasi-)circular, non-precessing orbit at large separations. In turn, the external gravitational field of the binary companion gives rise to the tidal field. The tidal environment is incorporated into the metric perturbation~$h_{\mu\nu}$ in Eq.~\eqref{eq:metric} based on a matched asymptotic expansion exploiting the hierarchy between the strong-field scale near the BH and the weak-field scale of the binary spacetime.

The leading-order description of the tidal environment in the weak-field zone is characterized by the gravitoelectric quadrupolar tidal moment, defined in terms of the frame components of the Weyl tensor~$C_{\alpha \mu \beta\nu}$, as follows~\cite{Poisson:2004cw,Poisson:2009qj}:
\begin{align}
    {\cal E}_{ij}\left(\lambda\right):= \left[C_{\alpha \mu \beta \nu}e^{\alpha}_0 e^\mu_i e^\beta_0 e^\nu_j\right]^{\rm STF}.
\end{align}
Here, $\lambda$ labels the worldline of the point mass approximately describing the deformed BH, $e^\mu_0$ is tangent to the worldline, $e^\mu_i$ forms an orthonormal triad, and ``STF'' denotes the symmetric and tracefree projection. For the binary systems considered in this work, the components of ${\cal E}_{ij}$ at leading order in the post-Newtonian (PN) expansion satisfy~\cite{Taylor:2008xy}
\begin{align}
    {\cal E}_{11}+{\cal E}_{22}=& -\frac{\epsilon}{M^2}\left[1+{\cal O}\left(M_{\rm T}/b\right)\right],\\
    {\cal E}_{11}-{\cal E}_{22}=& -\frac{3\epsilon}{M^2}\left[1+{\cal O}\left(M_{\rm T}/b\right)\right]\cos 2\Omega \lambda,\\
    {\cal E}_{12}=&-\frac{3\epsilon}{2M^2}\left[1+{\cal O}\left(M_{\rm T}/b\right)\right]\sin 2\Omega\lambda.
\end{align}
Here, $b$ is the binary separation, i.e., the distance between the Kerr BH and the perturber, $M_{\rm T}:=M+M_{\rm ext}$ is the total mass of the binary system together with the mass of the binary companion~$M_{\rm ext}$, where $\Omega:=\sqrt{M_{\rm T}/b^3}[1+{\cal O}(M_{\rm T}/b)]$ is the orbital angular velocity. The absence of ${\cal E}_{13}$ and ${\cal E}_{23}$ implies that the quadrupole tidal perturbation contains no $m=\pm1$ modes in its harmonic decomposition \cite{Taylor:2008xy}. The dimensionless, non-negative parameter~$\epsilon$ is defined by
\begin{align}
    \epsilon:= \frac{M^2 M_{\rm ext}}{b^3}.
    \label{eq:tidalamplitude}
\end{align}
The parameter $\epsilon$ controls, explicitly, the strength of the tidal field and, implicitly, the distance $b$ between the central object and the perturber, through Eq. \eqref{eq:tidalamplitude}. Henceforth, we ``christen'' $\epsilon$ the \emph{tidal amplitude}.

The validity of the PN description places an upper bound on $\epsilon$ such as~\cite{Katagiri:2026gkz}
\begin{align}
    \epsilon \lesssim \frac{1}{216}\times 
    \begin{cases}\dfrac{M_{\rm ext}}{M},\quad\qquad (M\ge M_{\rm ext}),\\
\left(\dfrac{M}{M_{\rm ext}}\right)^2,\quad (M\le M_{\rm ext}).\label{eq:PNrequirement}
    \end{cases}
\end{align}
For Sagittarius~A* with mass $M\sim 4 \times 10^6 M_\odot$, perturbed by a stellar-mass BH with mass $M_{\rm ext}\sim 40 M_\odot$ at a separation~$b\sim 5\, \textrm{AU}\simeq 127M$, i.e., the approximate distance between the Sun and Jupiter, the tidal amplitude~$\epsilon$ is $\mathcal{O}(10^{-12})$~\cite{Bronicki:2022eqa}. When the primary is a supermassive BH with $M=10^9M_\odot$ and is perturbed by another supermassive BH with $M=10^6 M_\odot$ at $b\sim 267\,{\rm AU} \simeq 27M$, which may occur in the late stages of galaxy mergers, the typical magnitude of $\epsilon$ is $\mathcal{O}(10^{-8})$ \cite{Bronicki:2022eqa}.

In the strong-field region near the deformed BH, the tidal perturbation is described in terms of the Weyl scalar~$\psi_0$, which is obtained as a solution of the Teukolsky equation in the stationary limit under the regularity condition at the BH horizon~\cite{LeTiec:2020bos,Katagiri:2026gkz}. By acting a linear operator, referred to as the Geroch-Held-Penrose operator~\cite{Geroch:1973am}, on $\psi_0$, the tidally-deformed metric~$h_{\mu\nu}$ is reconstructed. The normalization of the quadrupolar mode of $\psi_0$ is fixed by the tidal quadrupole moment through a scalar quantity~\cite{Katagiri:2026gkz}
\begin{align}
    &{\cal E}_{ij}n^i n^j=  \frac{\epsilon}{2M^2} \left[-1+3\cos^2\theta \right.\nonumber\\
    &\left. -3\cos\left(2\sqrt{\frac{M}{b^3}} \lambda-2\phi-2\Phi\right)\sin^2\theta\right]+{\cal O}\left(\epsilon^2\right).\label{eq:binarytidalmoment}
\end{align}
Here,~$n^i:=(\sin \theta \cos\varphi,\sin\theta \sin\varphi,\cos\theta)$ is the radial unit vector in spherical coordinates, where $\varphi$ denotes the azimuthal angle in the advanced Kerr coordinates. When rewriting it to $\phi$ in the Boyer-Lindquist coordinates through Eq.~\eqref{eq:FromaKtoBL}, we use $\varphi=\phi+\Phi$ with $\Phi:=(a/2\sqrt{M^2-a^2})\ln |(r-r_+)/(r-r_-)|$. Note that $\lambda$ should not be interpreted as a coordinate time. Rather, it labels an instantaneous snapshot of the geometry deformed by an effectively frozen tidal field. At fixed~$\lambda$, the tidal amplitude~$\epsilon$ introduced in Eq.~\eqref{eq:tidalamplitude} controls the amplitude of the instantaneous snapshot of $h_{\mu\nu}$. It follows from Eq.~\eqref{eq:binarytidalmoment} that the scalar spherical harmonic decomposition of the tidal scalar,~$\hat{\cal E}_{\ell m}$, contains no $m=\pm 1$ modes through ${\cal E}_{ij}n^i n^j=\hat{\cal E}_{2m}Y_{2m}$. 

One can choose $\lambda=0$ as a reference ``time''. This choice is equivalent to placing the binary companion at $(\theta,\phi)=(\pi/2,0)$ in the BH frame. As indicated by Eq.~\eqref{eq:binarytidalmoment}, the dependence of $h_{\mu\nu}$ on $\lambda$ and $\phi$ enters only through the phase in the form of $\sqrt{M/b^3}\lambda-\phi-\Phi$. A frozen configuration at $\lambda=0+\Delta \lambda$ is then obtained by the phase shift $\phi\to \phi-\sqrt{M/b^3} \Delta \lambda$ from the configuration at $\lambda=0$. Thus, varying $\lambda$ merely rotates the frozen tidal configuration. Note that corrections associated with the genuine binary evolution, such as radiation-reaction-driven changes in the binary separation, are beyond the stationary approximation considered here. Moreover, since the $\phi$ dependence appears through the phase~$2(\sqrt{M/b^3}\lambda-\phi-\Phi)$, the quadrupolar tidal deformation preserves a discrete symmetry under $\phi \to \phi+\pi$. Consequently, physically relevant orientations to the tidal field can be restricted to the range~$0\le \phi<\pi$.

In the present work, we assume $\epsilon\le {\cal O}(10^{-8})$. In geodesic motion, mainly considered in the remainder of this work, this setting corresponds to a weakly chaotic phase-space, where bound orbits can still occur. As $\epsilon$ increases, plunging or unbound orbits can appear before reaching the upper bound given in Eq.~\eqref{eq:PNrequirement}~\cite{Katagiri:2026gkz}. We emphasize that such small values of $\epsilon$ do not necessarily imply that the tidal effects are negligible because the components of $h_{\mu \nu}$ grow in $r$. Schematically, the tidal perturbation becomes comparable to the background at $r\sim (b^3/M_{\rm ext})^{1/2}$. Therefore, $h_{\mu\nu}$ should be understood as describing the local, strong-field geometry around the deformed BH, valid only up to $r\ll (b^3/M_{\rm ext})^{1/2}$, in the binary system~\cite{Poisson:2009qj}. 

\section{Geodesic motion and chaotic signatures}

The spacetime symmetries dictate that stationarity is a constant of motion while axisymmetry is (very mildly) broken; i.e., the test particle's energy $E$ is conserved, though the azimuthal angular momentum $L_z$ is not. The puzzle of symmetries is not completed with the existence of a Carter-like constant \cite{Ramond:2026bjy}. Thus, we expect phenomenology akin to chaos due to the imbalance of degrees of freedom between the spacetime and the secondary, test particle, of the EMRI. 

\subsection{Setup}

\subsubsection{System and initial conditions}

The geodesic equation can be written as
\begin{align}
\label{eq:GeodesicEq}
\,\,\ddot{x}^\kappa+\Gamma^\kappa_{\lambda\nu}\dot{x}^\lambda \dot{x}^\nu=0,
\end{align}
where $\Gamma^\kappa_{\lambda\nu}$ are the Christoffel symbols associated with the tidally-deformed spacetime, $\dot{x}^\kappa$ and $\ddot{x}^\kappa$ are the four-velocity and four-acceleration of the test particle, respectively, and the overdot denotes differentiation with respect to proper time. In the unperturbed Kerr case, the geodesic system~\eqref{eq:GeodesicEq} is integrable by virtue of the existence of three conserved quantities, i.e., the energy~$E:=-g_{\kappa\nu}(\partial/\partial t)^\kappa \dot{x}^\nu$, the azimuthal angular momentum~$L_z:=g_{\kappa\nu}(\partial/\partial \phi)^\kappa \dot{x}^\nu$, and the Carter constant \cite{Carter:1968rr}, together with the rest-mass conservation~$g_{\mu\nu} \dot{x}^\mu \dot{x}^\nu=-1$. The rest mass of the test-particle $\mu$ is set to unity for simplicity. In this scenario, bound geodesics exist in phase-space, described by their orbital elements, i.e., the eccentricity $e$, semi-latus rectum $p$, and inclination $\iota$. The corresponding fundamental frequencies of the orbit are defined by two libration frequencies; the periapsis-apoapsis oscillation frequency, $\omega_r$ and the polar oscillation through the equatorial plane, $\omega_\theta$. A final fundamental frequency, i.e., the revolution frequency $\omega_\phi$ completes the set. When all turning points of an orbit sit on top of the curve of zero velocity, as the rest-mass conservation defines, then they are space-filling, non-periodic, and form irrational ratios between the fundamental frequencies. Most orbits in Kerr are non-periodic or generic. Nevertheless, periodic or resonant orbits do exist in Kerr. These are not space-filling and return to their initial position after a number of librations take place. During resonance two or all fundamental frequencies are commensurable, depending on the nature of resonance. For example, orbital resonances in Kerr \cite{Flanagan:2010cd,Berry:2016bit} are defined in the context of commensurabilities with two fundamental frequencies, which occupy a zero-volume in the parameter space of orbits, while tidal resonances in tidally-deformed EMRIs \cite{Bonga:2019ycj} occur when all fundamental frequencies are in fractional unison.

Even though stationary tidal perturbations preserve time-translation symmetry, and hence $E$ remains conserved, they generically break axisymmetry and the Carter constant. Consequently, in the tidally-deformed spacetime, prolonged resonant motion of massive test particles can manifest \cite{Contopoulos_book}. This means that extended regions of commensurabilities, that span through a range of non-zero-volume parameter space conditions, can take place.
\begin{figure*}
\includegraphics[width=\textwidth]{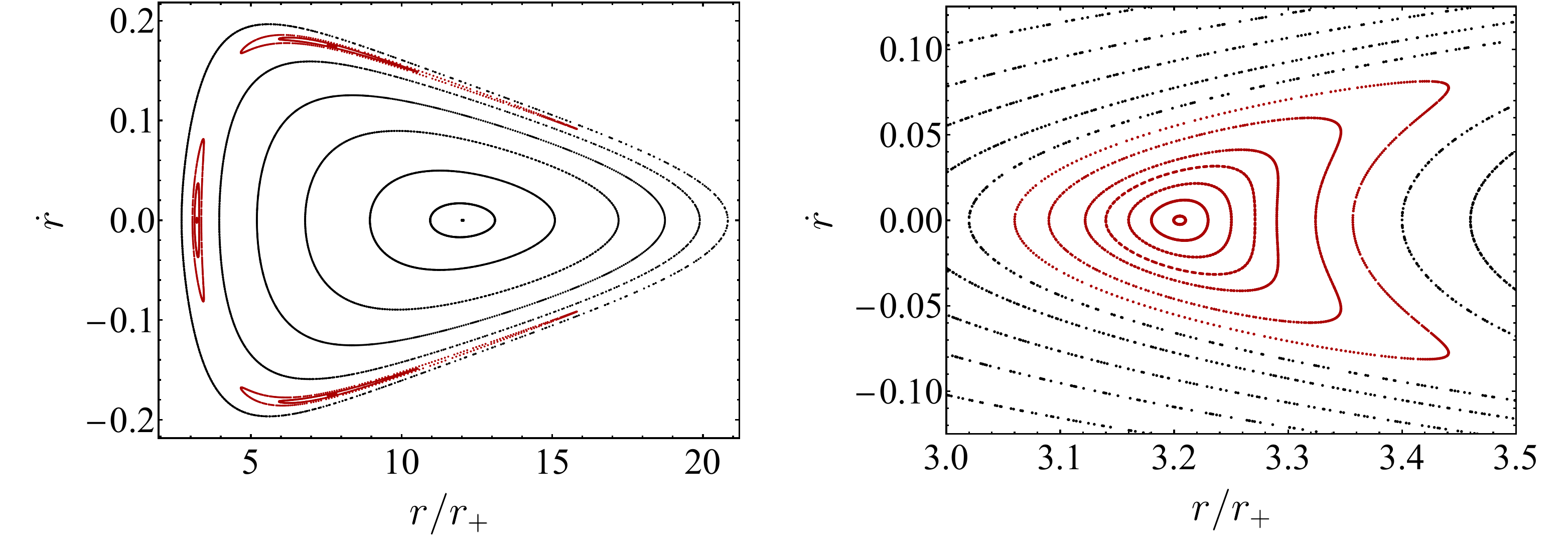}
\caption{\emph{Left:} Depiction of a Poincar\'e map at $\theta=\pi/2$ of a variety of orbits with fixed $E/\mu=0.98, \,L_z/\mu=1.8r_+$, where $\mu=1$ is the value of the massive test particle and $M$ is the mass of the tidally-deformed Schwarzschild BH. We have chosen the initial conditions of each orbit to be $(r[0]/r_+,\dot{r}[0]=0,\theta[0]=\pi/2,\dot{\theta}[0])$, where we vary $r[0]/r_+$, with $r_+=1$. The external tidal field has a tidal amplitude $\epsilon=10^{-8}$ and an initial tidal angle $\phi_{\rm tide}=0$. The demonstrated KAM curves are shown in black while the $2/3$-resonant island is shown with dark red color. \emph{Right:} Same as left but zoomed in on the strong-field region of one of the islands of stability, that is symmetric with respect to $\dot{r}=0$.}
\label{fig:PM_1}
\end{figure*}
\begin{figure}
 \includegraphics[width=\columnwidth]{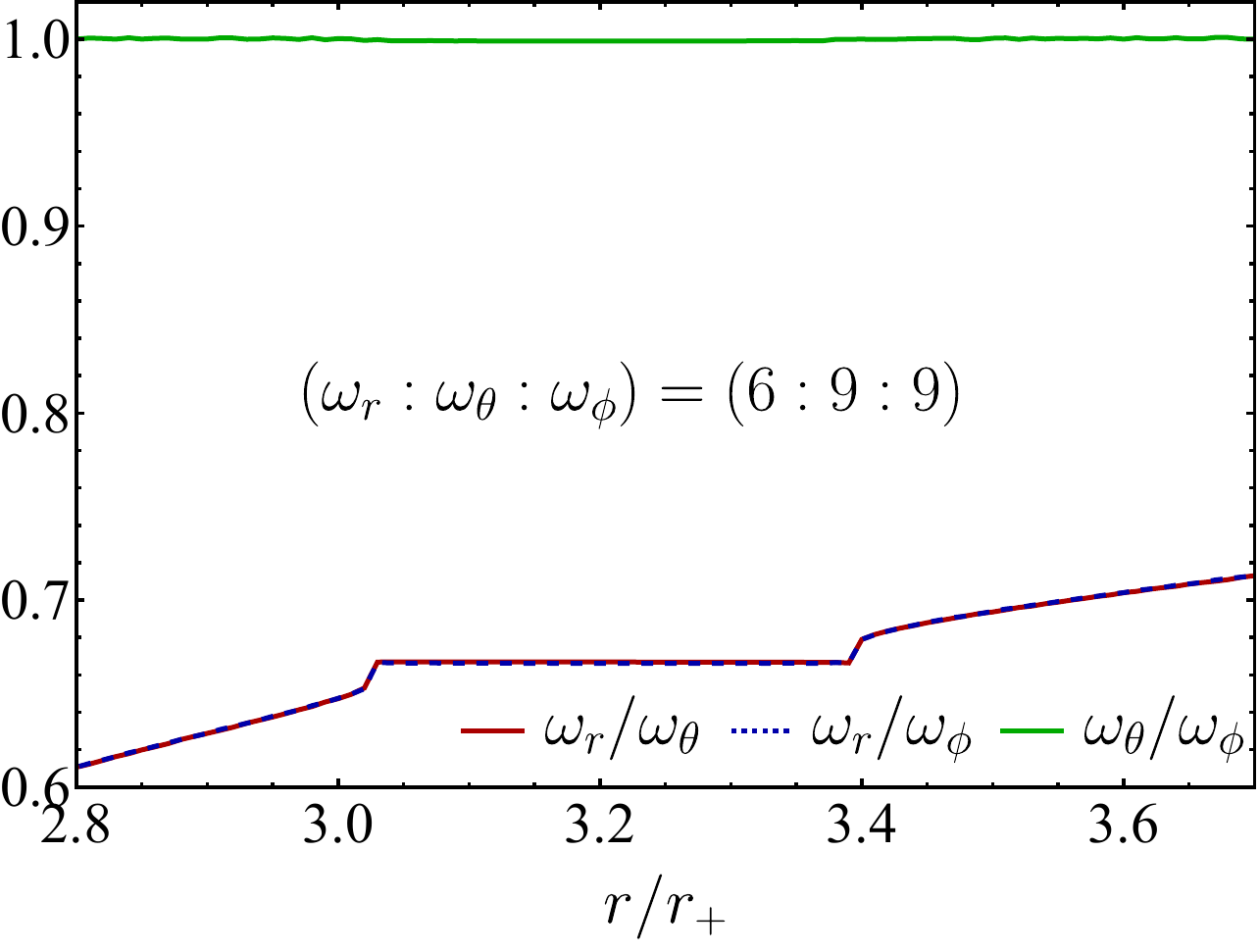}
    \caption{The fundamental frequency ratios, $\omega_r/\omega_\theta$, $\omega_r/\omega_\phi$, and $\omega_\theta/\omega_\phi$ for bound orbits that evolve with different initial radii in the tidally-deformed Schwarzschild spacetime. The tidal amplitude and angle are set to $\epsilon=10^{-8}$ and $\phi_{\rm tide}=0$. The secondary parameters are fixed as $E/\mu=0.98$ and $L_z/\mu=1.8r_+$.}
\label{fig:orbit_tide_res_Schwarz}
\end{figure}
\begin{figure}
\includegraphics[width=0.45\textwidth]{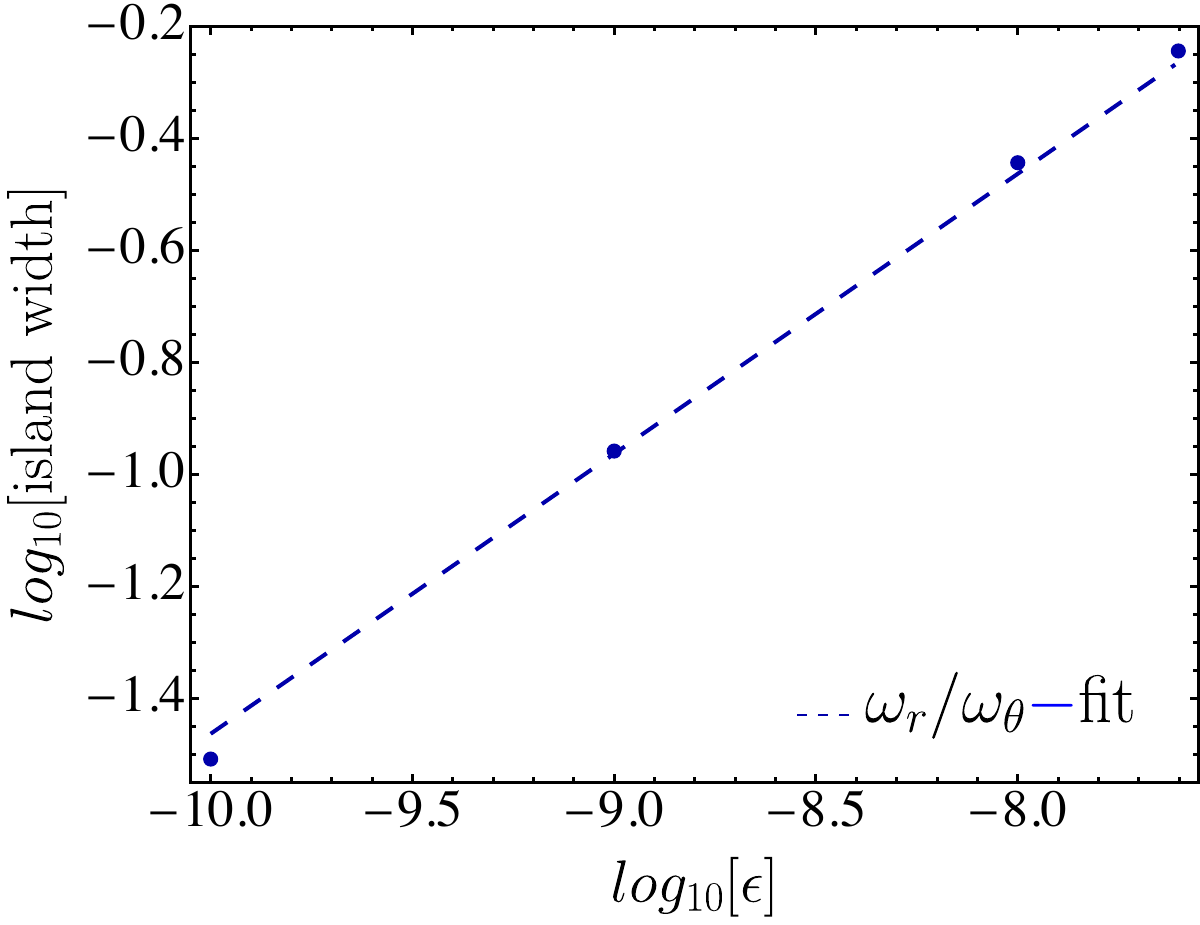}
\caption{$2/3$-resonant island width as a function of the tidal field $\epsilon$ for fixed $E/\mu=0.98,\,L_z/\mu=1.8r_+,\,\mu=1$ and $\phi_{\rm tide}=0, \, \dot{r}[0]=0$. The dots correspond to the island widths per $\epsilon$, while the dashed line is a linear fit of those points, with slope of $1/2$. The scaling is consistent with the fact that the resonance jump is proportional to the square root of the perturbation parameter.}
\label{fig:Schwarz_width_vs_eps}
\end{figure}
When implementing our geodesic solver (discussed below), we exploit the conservation of $E$ and $\mu$ for monitoring the accuracy of numerical integration of Eq.~\eqref{eq:GeodesicEq}. To evolve the system, we specify initial data at zero proper time, i.e., $\tau=0$. The zero proper time is considered as a reference point used to specify representative orbital configurations in an already tidally-deformed geometry. Different choices of an initial parameter set should be viewed as representative snapshots of possible EMRI configurations, rather than as modeling either the formation process or an instantaneous turn-on of the tidal perturbation. Specifically, we set $(r,\dot{r},\phi,\theta)|_{\tau=0}$\footnote{Or equivalently, $(r[0],\dot{r}[0],\phi[0],\theta[0])$.}, together with the conserved energy~$E$ and initial azimuthal angular momentum~$L_z|_{\tau=0}$. Although $L_z$ is no longer conserved, its initial value is used for fixing the initial azimuthal velocity. The remaining initial components are then determined from these constraints, together with the rest-mass conservation. Henceforth, we set $\theta|_{\tau=0}=\pi/2$. To consider several initial conditions to model different classes of representative configurations of tidally-deformed EMRIs, we choose various initial radii~$r|_{\tau=0}$. For each initial radius, we choose two initial radial momenta
\begin{align}
    \dot{r}|_{\tau=0}=\dot{r}[0]=0,0.1.
\end{align}
The former represents, in general, an instantaneous configuration of an eccentric orbit at a radial turning point. As a special case, this class includes a spherical orbit when $E$ and $L_z|_{\tau=0}$ are set appropriately. We also consider the case with finite initial radial momentum from a dynamical-system's point of view. A bound orbit passing through a point with $\dot{r}\neq0$ will generically pass through a radial turning point, and could therefore be included in orbits evolved from $\dot{r}=0$ with different orbital parameters. Our purpose is to check how resonant structures seen in the phase-space analysis below depend on the chosen phase-space slice \cite{Lukes-Gerakopoulos:2010ipp,Destounis:2021mqv}.

We, further, vary $\phi|_{\tau=0}$ to incorporate the effect of the relative orientation between the orbit and the effectively frozen tidal field in the BH frame. For clarity, we introduce the {\it tidal angle} defined by $\left.\phi_{\rm tide}:=\phi\right|_{\tau=0}$. As the simplest choices, we consider
\begin{align}
    \phi_{\rm tide}= 0, \,\pi/2, \,3\pi/2.
\end{align}
These cases correspond to the initial configuration aligned with or orthogonal to the direction towards the fixed tidal field, respectively. Although the system has a discrete symmetry under $\phi \to \phi + \pi$ as noted earlier, we explicitly consider both $\phi_{\rm tide}=\pi/2$ and $\phi_{\rm tide}=3\pi/2$ as a consistency check.

\subsubsection{Validity of approximations}

We now assess the validity of treating tidally-deformed EMRI evolution as geodesic motion on an effectively stationary geometry during bound motion of geodesics by comparing the relevant timescales. First, the timescale of variations of the evolving tidal field is estimated as \cite{Katagiri:2026gkz}
\begin{align}
    T_{\rm tides}\sim 2\pi \sqrt{\frac{M_{\rm ext}}{M+M_{\rm ext}}}\frac{M}{\sqrt{\epsilon}}.
\end{align} 
Then, the typical timescale of $n$ cycles of bound geodesics is
\begin{align}
    T_{\rm geo} \sim  2\pi n M.
\end{align}
Thus, for the geodesics during $n$ cycles, the geometry can be regarded as effectively stationary, provided that $T_{\rm tides} \gg T_{\rm geo}$, being equivalent to
\begin{align}
    n \ll \sqrt{\frac{M_{\rm ext}}{M+M_{\rm ext}}}\frac{1}{\sqrt{\epsilon}}.\label{eq:stationarity}
\end{align}
Next, we discuss the validity of the geodesic description. In the two-body problem of an unperturbed Kerr BH with secondary mass~$\mu$, the timescale of radiation-reaction-driven changes in the constants of motion is estimated as $T_{\rm rr}\sim M^2/\mu$. The geodesic approximation therefore remains valid over $n$ cycles if $T_{\rm rr} \gg T_{\rm geo}$, which gives
\begin{align}
     n \ll \frac{M}{\mu}.
\end{align}
Combining this with the stationarity condition~\eqref{eq:stationarity}, we obtain
\begin{align}
    1 \ll  n \ll {\rm min} \left[\sqrt{\frac{M_{\rm ext}}{M+M_{\rm ext}}}\frac{1}{\sqrt{\epsilon}},  \frac{M}{\mu}\right]\label{eq:conditionforn}
\end{align}
This hierarchy ensures that both the tidally-deformed geometry and $E$ are approximately fixed over many cycles. In what follows, our analysis focuses on bound orbits for which the number of cycles~$n$ satisfies Eq.~\eqref{eq:conditionforn}.

\subsection{Chaotic signatures}
%
\begin{figure}
\includegraphics[width=0.46\textwidth]{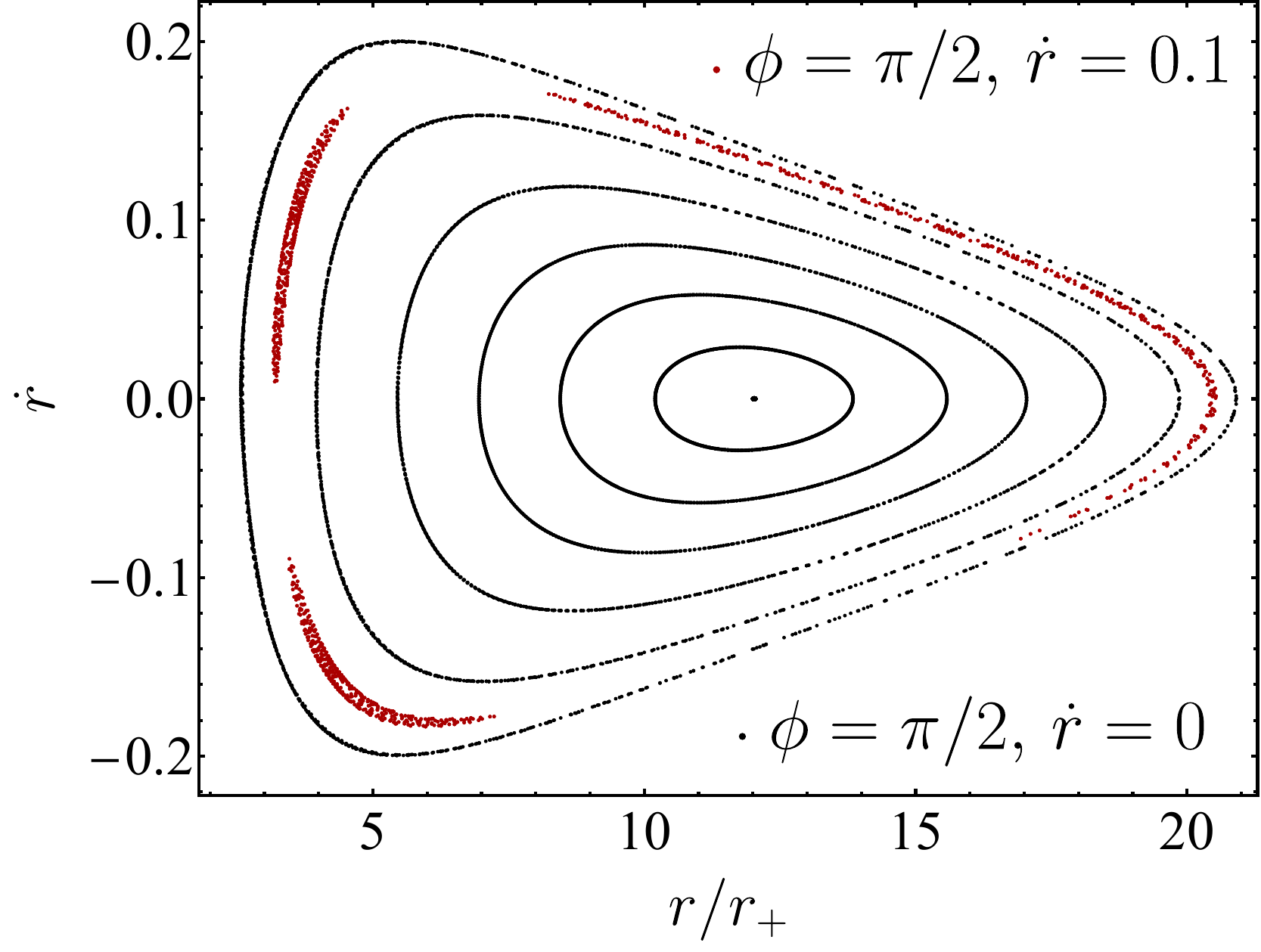}
\caption{Depiction of a Poincar\'e map at $\theta=\pi/2$ of a variety of orbits with fixed $E/\mu=0.98, \,L_z/\mu=1.8r_+$, where $\mu=1$ is the value of the massive test particle and $M$ is the mass of the tidally-deformed Schwarzschild BH. We have chosen the initial conditions of each orbit to be $(r[0]/r_+,\dot{r}[0]=0,\theta[0]=\pi/2,\dot{\theta}[0])$, where we vary $r[0]/r_+$, with $r_+=1$, for the black KAM curves. Due to the external tidal field, we have also set $\phi[0]=\pi/2$, $\dot{r}[0]=0.1$ with a tidal amplitude $\epsilon=10^{-8}$. The demonstrated KAM curves of the $2/3$-resonant island are shown with dark red color. We observe that if the initial condition of the radial momentum $\dot{r}[0]=0.1$, then a deformed Birkhoff chain emanates. On the other hand, an initial radial velocity of $\dot{r}=0$ does not reveal any Birkhoff chain since the horizontal line $\dot{r}=0$ does not intersect the deformed resonant islands from the left-hand side of the central fixed point.}
\label{fig:Schwarz_width_vs_phi}
\end{figure}
To decide if a spacetime's geodesics undergo chaotic phenomenology, we refer to the Kolmogorov-Arnold-Moser (KAM) \cite{Arnold_1963,Moser:430015} and the Poincar\'e-Birkhoff \cite{Birkhoff:1913} theorems. The former loosely states that if a small, non-integrable, perturbation is added to the Hamiltonian of an unperturbed dynamical system, then the KAM curves formed from the overall non-integrable system's orbits, slightly shift from the invariant integrable ones, but the overall map remains similar to an integrable one, as long as the geodesics are sufficiently far from resonances, such that they satisfy Arnold's criterion \cite{Arnold_1963}. The later theorem states that if an orbit is close to the vicinity of a resonance then the corresponding invariant curve breaks down into resonant islands of stability that encapsulate periodic stable orbits, and are finely surrounded by purely chaotic orbits that emanate from the unstable periodic orbits; a region in phase space that is called a Birkhoff chain (for a sketch of a Birkhoff chain, see Fig. 2 in \cite{Lukes-Gerakopoulos:2010ipp}). 

From a dynamical system point of view that regards EMRIs, chaos is extremely hard to spot directly; thus, one has to fine-tune the initial parameters of the orbit in order to find a vicinity of purely chaotic motion. On the other hand, there are certain indirect imprints of chaos in non-integrable EMRIs, i.e., the crossing of resonant islands. These phase space regions are equipped with an orbital periodicity that spans throughout the island, meaning that the ratio of the fundamental frequencies of any orbit inside the island, i.e.,  $\omega_r/\omega_\theta$, $\omega_r/\omega_\phi$, or $\omega_\theta/\omega_\phi$, is shared throughout it. Following earlier studies of non-integrable systems in axisymmetric spacetimes, we use $\omega_r/\omega_\theta$ to label resonant islands, although, in our non-axisymmetric system, $\omega_\phi$ plays an equally important role. Since, resonances in Kerr BHs, where integrability holds, are already an open issue in the EMRI modeling and data analysis \cite{Flanagan:2010cd,Ruangsri:2013hra,Berry:2016bit,Speri:2021psr,Levati:2025ybi,Eleni:2024fgs,Lynch:2024ohd,Levati:2026emw}, one can assume that breaking integrability should lead to stronger effects in the resulting waveforms of non-integrable EMRIs, thus making the issue of modeling resonances accurately even more complicated \cite{Dyson:2025dlj,Cardoso:2022whc,Destounis:2025tjn,Pan:2023wau}.
\begin{figure*}
\includegraphics[width=\textwidth]{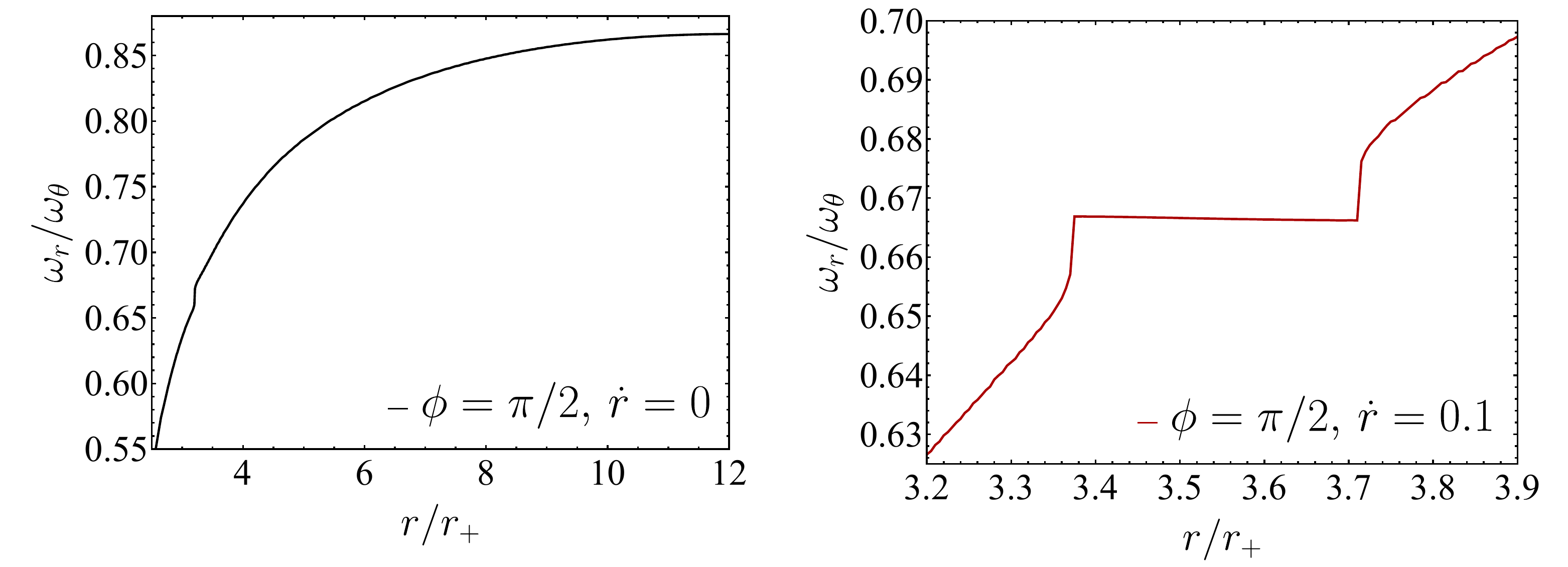}
\caption{\emph{Left:} Radial over polar frequency ratio (rotation curve) for a test particle with $\mu=1$ and $E/\mu=0.98,\,L_z/\mu=1.8r_+$ in tidally-perturbed Schwarzschild BH with the tidal amplitude $\epsilon=10^{-8}$ and tidal angle $\phi_{\rm tide}=\pi/2$. The test particle in study begins with $\dot{r}[0]=0$. We observe that due to the non-crossing of any resonant island (from the left-hand side of the central fixed point), but rather the crossing of the $\omega_r/\omega_\theta=2/3$ unstable periodic point region in the Birkhoff chain, we only see an inflection point in the rotation curve; a typical manifestation of unstable periodic orbits that give rise to chaotic layers. \emph{Right:} Same as left but with an initial radial momentum $\dot{r}[0]=0.1$. This initial velocity drives the test particle through the $2/3$-resonant island (see Fig. \ref{fig:Schwarz_width_vs_phi}), which manifests through a plateau in the rotation curve.}
\label{fig:Schwarz_width_vs_phi_nonzero}
\end{figure*}
%
\subsection{Non-rotating case}
 
In what follows, we study how bound geodesics around a tidally-deformed 
Schwarzschild BH respond to the change in the tidal amplitude~$\epsilon$ and the tidal angle~$\phi_{\rm tide}$.
 
In Fig. \ref{fig:PM_1} we demonstrate the effect of a weak tidal field at $\phi_{\rm tide}=0$ with $\epsilon=10^{-8}$. Regardless of the small amplitude, we observe that the Poincar\'e map shows signs of non-integrability due to the formation of $2/3$-resonant islands, shown with dark red. These islands present an indirect signature of chaos, since inside the resonant island the ratio $\omega_r/\omega_\theta=2/3$ is preserved, while the bordering vicinity is covered with chaotic orbits (not visible in Fig.~\ref{fig:PM_1}). Indeed, Fig.~\ref{fig:orbit_tide_res_Schwarz} shows that a very clear plateau appears at $\omega_r/\omega_\theta=\omega_r/\omega_\phi=2/3$ which directly renders the system non-integrable, due to the tidal perturbation. The respective plateau has the same width as the one of the resonant island. On the other hand, the small tidal amplitude does not significantly interfere with the Schwarzschild BH symmetries so that the ratio $\omega_\theta/\omega_\phi\simeq 1$ for all the scanned parameter space, although the weak tidal deformation, in principle, splits $\omega_\theta$ and $\omega_\phi$. In Section~\ref{sec:tidalresonance}, we discuss resonant dynamics suggested by the plateaus in terms of action-angle descriptions.
 
To comprehend the relation of the $2/3$-island widths with the tidal amplitude~$\epsilon$, we show in Fig.~\ref{fig:Schwarz_width_vs_eps} that the plot of the island width versus $\epsilon$ forms a straight line (in the log-log scale), which coincides with Schwarzschild EMRIs with rotating secondaries (see Fig. 7 in \cite{Zelenka:2019nyp}) and charged, magnetized EMRI analogs (see Fig. 3 in \cite{Mukherjee:2022dju}). We will see that the same does not hold for tidally-perturbed Kerr EMRIs due to the orbit-tide coupling.

Figure~\ref{fig:Schwarz_width_vs_phi} shows 
that different choices of $\phi_{\rm tide}$ give rise to a noticeable impact on the Poincar\'e maps and respective rotation curves. Specifically, asymmetric Birkhoff islands form with respect to the axis~$\dot{r}=0$, when $\phi_{\rm tide}=\pi/2$. This asymmetric structure can be traced to the absence of an azimuthal Killing vector. To see this explicitly, we consider the Lagrangian of the secondary, ${\cal L}=(1/2)g_{\kappa\nu} \dot{x}^\kappa \dot{x}^\nu$. The radial Euler-Lagrange equation at $\theta = \pi/2$ leads to
\begin{align}
    g_{rr} \ddot{r}=\frac{1}{2}\frac{\partial g_{tt}}{\partial r} \dot{t}^2-\frac{1}{2}\frac{\partial g_{rr}}{\partial r}\dot{r}^2+\frac{1}{2}\frac{\partial g_{\phi \phi}}{\partial r}\dot{\phi}^2-\frac{\partial g_{rr}}{\partial \phi} \dot{r}\dot{\phi}.
\end{align}
The last term presents a coupling between radial and polar velocities, $\dot{r}\dot{\phi}$, and arises because the tidally-deformed geometry depends on $\phi$, explicitly. This is an unambiguous sign of orbit-tide coupling. The emergence of this nonlinear term implies that the geodesic equations are not invariant under $\dot{r} \to - \dot{r}$. This suggests that the set of intersection points in the Poincar\'e map, $(r,\dot{r})|_{\theta=\pi/2}$, is not necessarily symmetric with respect to the horizontal axis $\dot{r}=0$. Similar behavior has been found in non-circular, beyond-Kerr BHs \cite{Zhou:2021cef,Chen:2023gwm}, where asymmetric Birkhoff islands form with respect to those generated in typical non-integrable, but circular spacetimes.

Finally, in the left panel of Fig.~\ref{fig:Schwarz_width_vs_phi_nonzero}, we show that the rotation curve possesses an inflection point around $2/3$, since the choice $\phi_{\rm tide}[0]=\pi/2, \, \dot{r}[0]=0$ leads to a passage close to the vicinity of an unstable periodic orbit. When we choose $\phi_{\rm tide}[0]=\pi/2, \, \dot{r}[0]=0.1$,  a well-defined plateau is demonstrated in the right panel of Fig. \ref{fig:Schwarz_width_vs_phi_nonzero}. This plateau corresponds to the crossing of the top left lopsided island in Fig. \ref{fig:Schwarz_width_vs_phi}.

\subsection{Rotating case}
%
\begin{figure*}
\includegraphics[width=\textwidth]{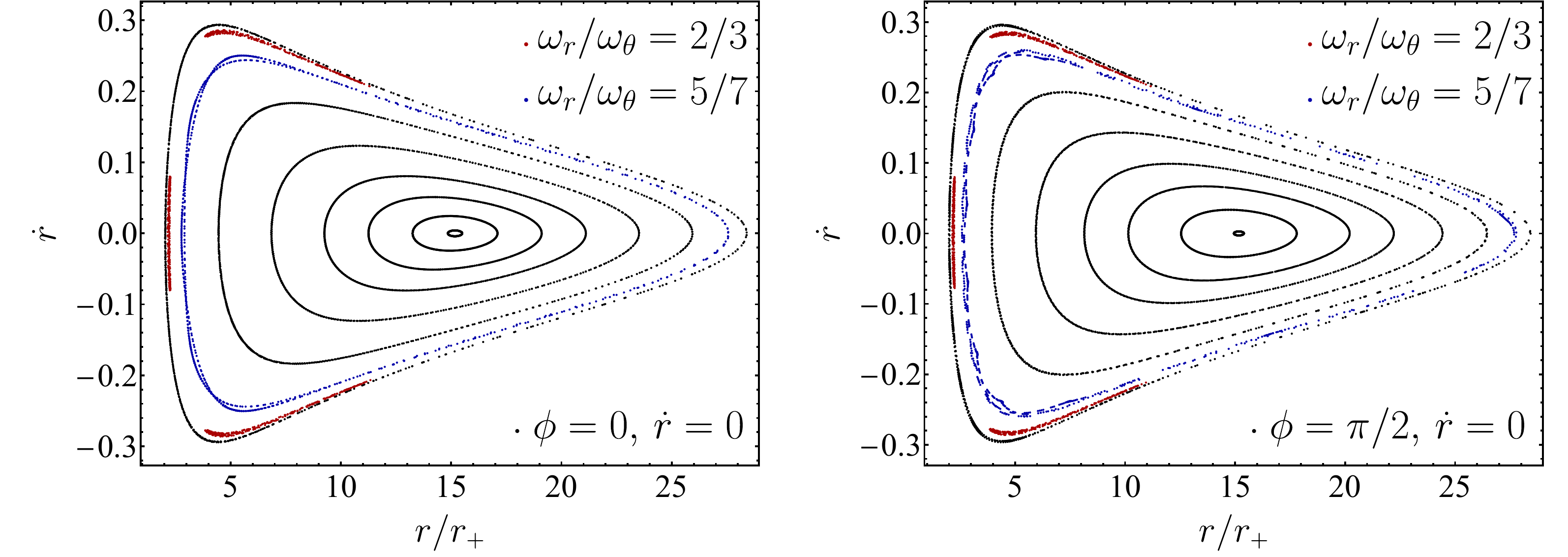}
\caption{\emph{Left:} Poincar\'e map of a tidally-deformed Kerr BH with $a=0.8M$. The tidal amplitude is $\epsilon=10^{-8}$. The test particle has fixed $E/\mu=0.98,\,L_z/\mu=1.8r_+$, while its initial conditions are set to $(r[0]/r_+,\dot{r}[0]=0,\theta[0]=\pi/2,\dot{\theta}[0])$, where $\dot{\theta}[0]$ is defined by the constraint equations. In this case, we show the KAM curves with black dots, and the islands $2/3$ and $5/7$ with dark red and dark blue, respectively, where $\phi=0$ and $\dot{r}[0]=0$. \emph{Right:} Same as left with $\phi_{\rm tide}=\pi/2$ and $\dot{r}[0]=0$. We observe that in contrast to the tidally-perturbed Schwarzschild, the $2/3$ and $5/7$ Birkhoff chains are distorted accordingly so that an island crossing is still possible when $\phi_{\rm tide}=\pi/2$, in contrast to what occurs in a tidally-deformed Schwarzschild spacetime.}
\label{fig:Kerr_maps_vs_phi}
\end{figure*}
\begin{figure}
\includegraphics[width=\columnwidth]{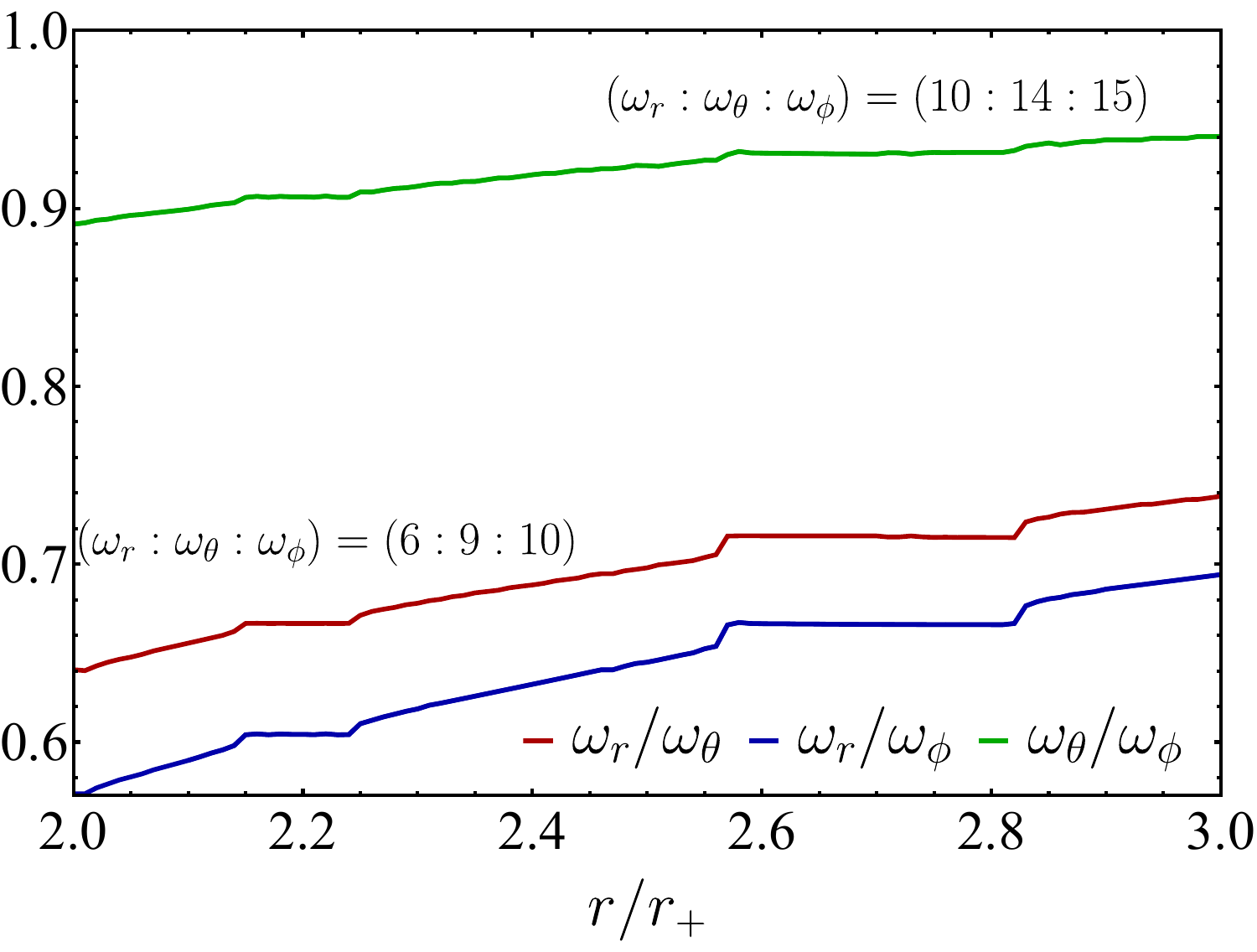}
    \caption{The fundamental frequency ratios, $\omega_r/\omega_\theta$, $\omega_r/\omega_\phi$, and $\omega_\theta/\omega_\phi$ for bound orbits that evolve with different initial radii in a tidally-deformed Kerr spacetime with spin $a=0.8M$. The tidal amplitude and angle are set to $\epsilon=10^{-8}$ and $\phi_{\rm tide}=0$. The secondary parameters are fixed as $E/\mu=0.98$ and $L_z/\mu=1.8r_+$. The first column of plateaus correspond to the triplet $(\omega_r:\omega_\theta:\omega_\phi)=(6:9:10)$, while the second one corresponds to $(\omega_r:\omega_\theta:\omega_\phi)=(10:14:15)$.}
    \label{fig:orbit_tide_res_Kerr}
\end{figure}
\begin{figure*}
\includegraphics[width=\textwidth]{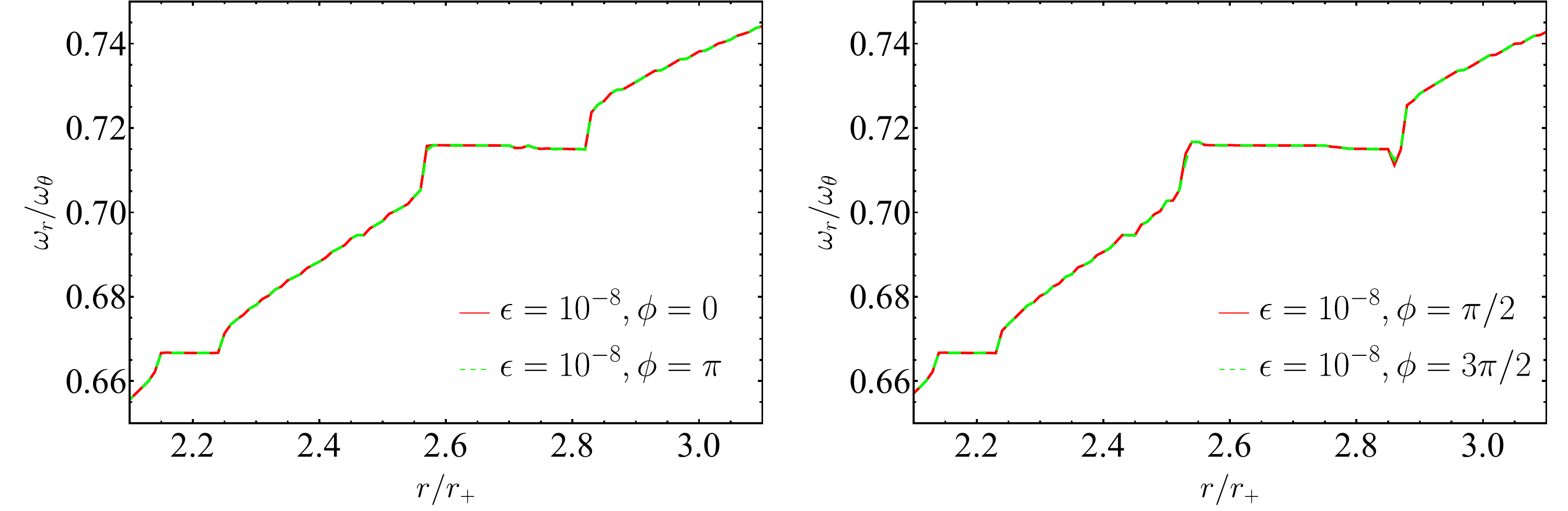}
\caption{Frequency ratios $\omega_r/\omega_\theta$ of a test particle orbiting a tidally-perturbed Kerr BH, with spin $a=0.8M$ and tidal field amplitude $\epsilon=10^{-8}$, for various initial locations $r[0]/r_+$ and angles $\phi_{\rm tide}=0,\,\pi/2,\,\pi,\,3\pi/2$. We have chosen $\dot{r}[0]=0$, $\theta=\pi/2$, while the remaining initial momenta for $\theta$ and $\phi$ are defined by the constraint equations.}
\label{fig:rotcurvesphi}
\end{figure*}
\begin{figure}
\includegraphics[width=0.48\textwidth]{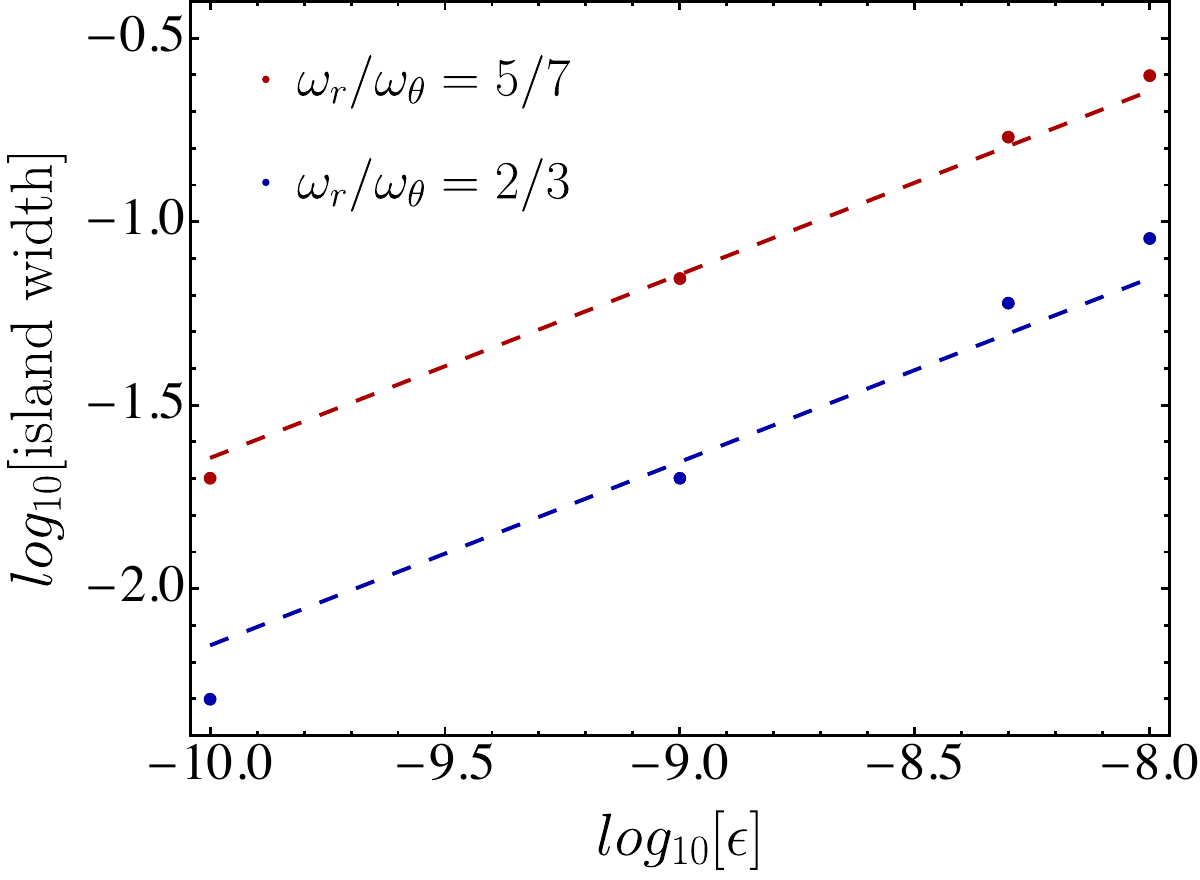}
\caption{Resonant island width of the first two frequency ratios $\omega_r/\omega_\theta$ of a test particle orbiting a tidally-perturbed Kerr BH, with spin $a=0.8M$ and varying tidal field amplitude $\epsilon$. We have fitted the widths corresponding to both of these resonances with a straight line of slope 1/2, as in the earlier non-rotating case.}
\label{fig:widthvseps}
\end{figure}
In this section, we promote the previous analysis to a tidally-perturbed Kerr supermassive BH, in order to understand the effect of spin on the tidal phenomenology, thus including its effects in the relativistic, restricted, three-body problem.

We commence the analysis with the Poincar\'e maps shown in Fig. \ref{fig:Kerr_maps_vs_phi}. The maps (also shown in Ref. \cite{Katagiri:2026gkz} for varying $\epsilon$ and fixed $\phi_{\rm tide}=0$) demonstrate a more complicated structure of resonant islands and non-integrability. In fact, besides the typical KAM curves of the $2/3$ island, another island set appears, i.e., the $5/7$ island. The initial velocity of the test particle is fixed to $\dot{r}[0]=0$, while we choose two cases of the tidal field's azimuthal position, namely, $\phi_{\rm tide}=0, \pi/2$.

Figure~\ref{fig:orbit_tide_res_Kerr} demonstrates the rotation curves associated with the Poincar\'e maps in Fig. \ref{fig:Kerr_maps_vs_phi}. We find that the ratios $\omega_r/\omega_\theta$~(dark red) and $\omega_r/\omega_\phi$~(dark blue) exhibit plateaus at $2/3$ and $5/7$. In contrast to the non-rotating case, $\omega_\theta$ and $\omega_\phi$ are clearly distinct, reflecting the absence of spherical symmetry in the unperturbed Kerr limit. Consequently, $\omega_\theta/\omega_\phi$~(dark green) also exhibits plateaus at $9/10$ and $14/15$, over the same range of radii as in $\omega_r/\omega_\theta$ and $\omega_r/\omega_\phi$. In Section~\ref{sec:tidalresonance}, we will deepen our understanding of these effects through the evolution of geodesics inside and outside the plateaus, in action-angle variables.

A first observation is the fact that due to the plurality and multiplicity of the islands ($2/3$ and $5/7$), the dependence on $\phi_{\rm tide}$ is more intricate than that observed in tidally-deformed Schwarzschild EMRIs, and it is, surprisingly, focused on the $5/7$ island (the $2/3$ island is symmetric to a 90 degree rotation of $\phi_{\rm tide}$). We expect that this will give a different orbital phenomenology than that of Fig. \ref{fig:PM_1} and \ref{fig:Schwarz_width_vs_phi}. Indeed, in Fig. \ref{fig:rotcurvesphi}, we see that two plateaus appear for all chosen $\phi_{\rm tide}$ and fixed $\epsilon$. 

The symmetries of the rotation curves hold under a 180-degree rotation to the initial tide position; an expected result due to the sinusoid dependence of spacetime with respect to $\phi$. This is also observed in Fig. \ref{fig:Schwarz_width_vs_phi}, where the dependence of the island widths is indeed periodic, with periodicity $\pi$. Scanning through the azimuthal angle of the tidal field, we find a preference of the system to give wider plateaus for $\phi_{\rm tide}=\pi/2, 3\pi/2$. The percentage difference between the $\phi_{\rm tide}=0, \pi/2$ is $\sim18\%$, for the $5/7$ plateau, while for the $2/3$ it is $\sim 2\%$.

The dependence on the tidal amplitude $\epsilon$ has a similar behavior as that found for the non-rotating tidally-deformed spacetime. The increment of $\epsilon$ leads to an expected increment to the island widths, which we demonstrate in Fig. \ref{fig:widthvseps}, where the $\epsilon$ dependence on the island width seems polynomial. 

\section{Tidal resonances in EMRIs}
\label{sec:tidalresonance}

In EMRIs, the presence of an external tidal perturber can induce secular changes in the constants of motion, causing measurable shifts in the accumulated GW phase~\cite{Bonga:2019ycj,Gupta:2021cno}. This phenomenon is known as a tidal resonance and occurs when the fundamental frequencies satisfy the resonance condition
\begin{align}
n \omega_r+k \omega_\theta+m \omega_\phi=0,\label{eq:TidalResonanceCondition}
\end{align}
with $n,k,m\in \mathbb{Z}$. This is the same commensurability condition underlying the finite-width plateaus emerging in the rotation curve for geodesic motion in an effectively frozen tidally-deformed BH spacetime. In what follows, we will examine the connection between tidal resonances and the plateaus observed in the rotation curves in Figs.~\ref{fig:orbit_tide_res_Schwarz} and~\ref{fig:orbit_tide_res_Kerr} in terms of the action-angle description \cite{Schmidt:2002qk}.

Within the present setup, i.e., a non-precessing, (quasi-) circular binary, the quadrupolar tidal perturbation contains only the modes with $m=0$ and $m=\pm2$. The resonance condition~\eqref{eq:TidalResonanceCondition} therefore reduces to the following two families:
\begin{align}
n \omega_r+k \omega_\theta=&0,\quad n \omega_r+
k \omega_\theta\pm 2 \omega_\phi=0.\label{eq:TidalResonanceConditionm0pm2}
\end{align}
For the non-rotating case, Fig.~\ref{fig:orbit_tide_res_Schwarz} implies $(\omega_r:\omega_\theta:\omega_\phi)=(6:9:9)$. Together with the resonance condition~\eqref{eq:TidalResonanceConditionm0pm2}, these ratios suggest the following possible resonant integer combinations
\begin{align}
    (n,k,m)=\pm(3,-2,0), \pm(3,0,-2),\label{eq:tripletsfornonrotating}
\end{align}
and their non-zero integer multiples. Note that these resonance conditions are nearly identical because $\omega_\theta \simeq \omega_\phi$. As shown in the next section, the angle-variable combinations specified by these integers are responsible for tidal resonances. For the rotating case, Fig.~\ref{fig:orbit_tide_res_Kerr} shows that the first plateaus, $(\omega_r:\omega_\theta:\omega_\phi)=(6:9:10)$ satisfy the resonance condition~\eqref{eq:TidalResonanceConditionm0pm2} for
\begin{align}
(n,k,m)=\pm (3,-2,0),\label{eq:lowertripletsforrotating}
\end{align}
while, at the second plateau, $(\omega_r:\omega_\theta:\omega_\phi)=(10:14:15)$, i.e., 
\begin{align}
(n,k,m)=\pm(3,0,-2),\pm(7,-5,0).\label{eq:highertripletsforrotating}
\end{align}
The next section shows that the angle combination associated with $\pm(3,0,-2)$ drives tidal resonances in rotating tidally-deformed EMRIs.

\subsection{Resonances in action-angle variables}

Our analysis in the previous section focused on the evolution of the physical coordinates~$x^\mu=(t,r,\theta,\phi)$ and their conjugate momenta~$p_\mu$ as functions of proper time~$\tau$. This description is particularly useful for identifying chaotic signatures and the resonant parameter space in phase space. We now turn to an action-angle description, which is commonly used to characterize the evolution of nearly integrable orbits under weak perturbations, such as the gravitational self-force~\cite{Flanagan:2010cd,Hinderer:2008dm} or an external tidal perturbation~\cite{Bonga:2019ycj,Gupta:2021cno}. For the unperturbed Kerr system, the action-angle variables~$(q^\alpha,J_\alpha)$ are obtained from $(x^\mu,p_\mu)$ through a canonical transformation~\cite{Schmidt:2002qk}, as outlined below.

In general, secondary motion in EMRIs subject to an external tidal field deviates from geodesic motion in the Kerr spacetime because of both self-force and tidal effects. The evolution of the angle~$q^\alpha$ and action variables~$J_\alpha$ can be described perturbatively as~\cite{Hinderer:2008dm,Bonga:2019ycj}
\begin{align}
\frac{dq^\alpha}{d\tau}=&\omega^\alpha_{\rm Kerr}+\epsilon g_{{\rm tides}}^{\alpha,(1)}+\eta g_{{\rm sf}}^{\alpha,(1)}+{\cal O}\left(\eta^2,\epsilon^2,\epsilon\eta \right),\label{eq:dotqalpha}\\
    \frac{d J_\alpha}{d\tau}=&\epsilon G_{\alpha,{\rm tides}}^{(1)}+\eta G_{\alpha,{\rm sf}}^{(1)}+{\cal O}\left(\epsilon^2,\eta^2,\epsilon \eta\right),\label{eq:dotJalpha}
\end{align}
where $\omega^\alpha_{\rm Kerr}=\omega^\alpha_{\rm Kerr}(J_\alpha)$ is a set of fundamental frequencies of geodesic motion in the unperturbed Kerr spacetime, and $\eta=\mu/M$ is the mass ratio. Here, the tidal and self-force contributions are specified by ``tides'' and ``sf'' subscripts with the small parameters~$\epsilon$ and $\eta$, respectively. Henceforth, we focus only on the tidal contributions. 

Before continuing to the explanation of the tidal contributions to the action-angle variables, we briefly recall a basic feature of oscillatory motion in celestial mechanics in terms of an effective pendulum description. Specifically, we introduce the notion of phase locking that will play a central role in interpreting resonances in this section. The angles, $q^r,q^\theta,q^\phi$, specify the phase of radial, polar, and azimuthal motions. Phase locking occurs when two or more oscillatory motions become synchronized so that their relative timing does not steadily slip, often due to physical interaction or coupling between them. The two possible behaviors of the relative phase near a resonance are described by either libration or circulation. In a phase-locked state, the phase librates: it is confined to a finite interval, e.g., $-\pi<\psi<\pi$, for the phase~$\psi$. In an effective pendulum description, the corresponding trajectory oscillates around a stable equilibrium point and lies inside the separatrix, or equivalently has an energy below the separatrix energy. An example in celestial mechanics is the Earth-Moon system which is tidally phase-locked. Due to the eccentric orbit of the Moon, it undergoes libration, meaning that it appears to gently wobble from the Earth's perspective, allowing us to see about $59\%$ of its surface over time. In a non-phase-locked state, the phase circulates: it continuously increases or decreases and repeatedly pass through a full $2\pi$ cycle. A simple analogy here is a pendulum swinging fast enough to loop over the top repeatedly, and circulation corresponds to motion above the separatrix energy. For more information regarding the distinction between libration and circulation in solar system dynamics and, in particular, the restricted three-body problem we refer the reader to Refs.~\cite{Murray:1999ssd..book.....M,Miller:1991BAAS...23.1314M}.

Equation~\eqref{eq:dotJalpha} shows that $G_{\alpha, {\rm tides}}^{(1)}$ is the leading-order contribution to changes in $J_\alpha$. At tidal resonances, resonant orbits lead to secular changes in $J_\alpha$, producing accumulated GW dephasing over many orbital cycles. To understand this accumulated effect, we expand  $G_{\alpha, {\rm tides}}^{(1)}$ in a Fourier series in the angle variables: 
\begin{align}
    G_{\alpha, {\rm tides}}^{(1)}\left(q^\alpha,J_\alpha \right)=\sum_{n,k,m}G_{\alpha,nkm}\left(J_\alpha\right)e^{i \left(n q^r+k q^\theta+m q^\phi\right)}.\label{eq:GtidalinFourier}
\end{align}
As shown below, for a non-resonant orbit, the phase~$n q^r+k q^\theta+m q^\phi$ generically circulates rapidly. Consequently, the corresponding~$(n,k,m)$ Fourier component oscillates rapidly and hence is zero in time-averaging over many orbital cycles. By contrast, for a resonant orbit satisfying Eq.~\eqref{eq:TidalResonanceCondition}, the corresponding phase is generically locked and evolves slowly. The corresponding component can then persist after time-averaging and thus produce secular changes in $J_\alpha$. This is exactly the underlying mechanism of tidal resonances in EMRIs.

Here, we study the evolution of the phase~$n q^r+k q^\theta+ m q^\phi$ associated with tidal resonances. For later convenience, for a particular resonance specified by integers~$(n_{\rm res},k_{\rm res},m_{\rm res})$, we define the corresponding resonance angle combination by
\begin{align}
    \psi= n_{\rm res} q^r+k_{\rm res} q^\theta+m_{\rm res} q^\phi.\label{eq:psi}
\end{align}
It is useful to discuss the timescale of changes in $\psi$ in terms of its derivative with respect to proper time~$\tau$:
\begin{align}
    \frac{d\psi}{d\tau}=&n_{\rm res} \frac{d q^r}{d\tau}+k_{\rm res} \frac{d q^\theta}{d\tau}+m_{\rm res} \frac{d q^\phi}{d\tau},\nonumber\\
    \simeq & n_{\rm res} \omega_r^{\rm Kerr}+k_{\rm res} \omega_\theta^{\rm Kerr}+m_{\rm res} \omega_\phi^{\rm Kerr}+\epsilon \Delta \Omega_{\rm tides}^{(1)},\label{eq:psidot}
\end{align}
where $\Delta \Omega_{\rm tides}^{(1)}$ arises from the combination of $g_{{\rm tides}}^{\alpha,(1)}$ in Eq.~\eqref{eq:dotqalpha}. The second line shows that, since near the resonance specified by $(n_{\rm res},k_{\rm res},m_{\rm res})$, the zeroth-order frequency combination is small, ${d\psi/d\tau}$ is at most ${\cal O}(\epsilon)$. Consequently, $\psi$ varies in time slowly and generically librates, as we will show below. By contrast, for non-resonant orbits, the zeroth-order frequency combination in Eq.~\eqref{eq:psidot} dominates over the tidal contribution and remains nearly constant. This indicates the circulation of $\psi$ with time. 

We begin by introducing key equations in the action-angle approach for geodesic motion in the unperturbed Kerr spacetime. The Hamiltonian for a particle along a bound geodesic orbit in the unperturbed Kerr spacetime is defined by~\cite{Carter:1968rr}
\begin{align}
{\cal H}=\frac{1}{2}(g^{\rm Kerr})^{\mu \nu} p_\mu p_\nu.
\end{align}
The conserved value of ${\cal H}$ is $-\mu^2/2$. The momenta~$p_\mu$ can be expressed in terms of the Boyer-Lindquist coordinates~$x^\mu$ and the constants of motion as\footnote{There is a sign uncertainty in $p_r$ and $p_\theta$. Here, we choose the positive branch for both momenta.}
\begin{align}
    p_t=-E,~p_r=\frac{\sqrt{V_r}}{\Delta},~p_\theta=\sqrt{V_\theta},~p_\phi=L_z.
\end{align}
where the two potentials read
\begin{align}
  V_r=&\left[\left(r^2+a^2\right)^2E-aL_z\right]^2-\Delta \left[\mu^2r^2+\left(L_z-a E\right)^2+Q\right],\\
  V_\theta=&Q-\left[\left(\mu^2-E^2\right)a^2+\frac{L_z^2}{\sin^2\theta}\right]\cos^2\theta,
\end{align}
Here, $Q$ is the Carter constant,
\begin{align}
    Q=p_\theta^2+a^2\cos^2\theta \left(\mu^2-p_t^2\right)+\cot^2\theta p_\phi^2.
\end{align}
\begin{figure*}[t]
\includegraphics[width=\textwidth]{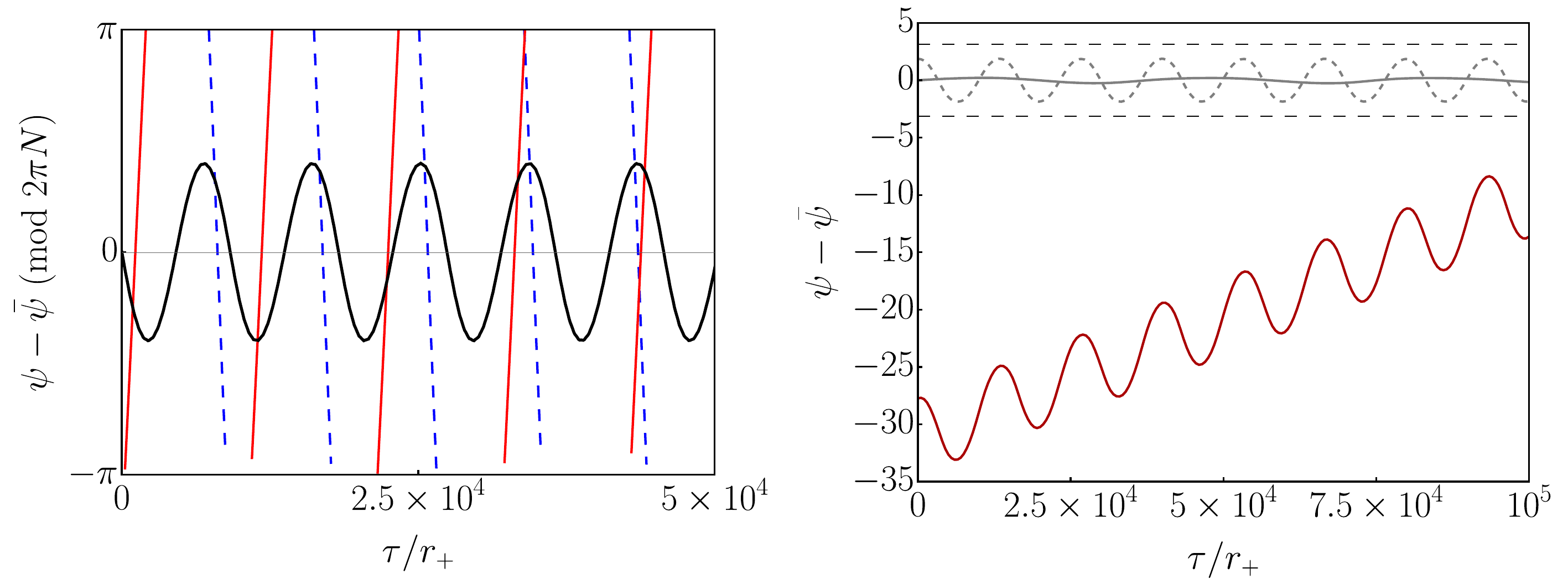}
    \caption{{\it Left:} Evolution of $\psi$ for the resonance~$(n_{\rm res}, k_{\rm res},m_{\rm res})=(3,-2,0)$ in the non-rotating case with the same orbital parameters as in Fig.~\ref{fig:orbit_tide_res_Schwarz}. Here, $\bar{\psi}$ is the mean value of $\psi$. The dashed-blue and red lines correspond to orbits evolved from $r/r_+=2.8, 3.8$, respectively, i.e., outside the plateau shown in Fig.~\ref{fig:orbit_tide_res_Schwarz}. The black curve represents an orbit evolved from $r/r_+=3.1$, inside the same plateau. For presentation purposes, we choose $N=6$. {\it Right:} Evolution of $\psi$ in the rotating case with $a=0.8M$, for orbits with initial radii within the plateaus shown in Fig.~\ref{fig:orbit_tide_res_Kerr}. The solid gray curve shows the libration of $\psi$ with $(n_{\rm res},k_{\rm res},m_{\rm res})=(3,-2,0)$ along an orbit evolved from $r/r_+=2.2$. The dashed gray and solid dark red curves show $\psi$ for $(3,0,-2)$ and $(7,-5,0)$, respectively, along the same orbit evolved from $r/r_+=2.7$. The $(3,0,-2)$ resonant angle librates in the range between $-\pi$ and $\pi$, marked by the horizontal black dashed lines, whereas the $(7,-5,0)$ resonant angle circulates with oscillations.}
    \label{fig:ResonantAngles}
\end{figure*}
The action variables~$J_\alpha$ for a bound geodesic in the unperturbed Kerr spacetime read~\cite{Schmidt:2002qk,Hinderer:2008dm}
\begin{equation}
\begin{aligned}
    &J_t=\frac{1}{2\pi} \int_0^{2\pi} dt~p_t,~J_r=\frac{1}{2\pi}\oint dr~p_r,\\
    &J_\theta=\frac{1}{2\pi}\oint d\theta~p_\theta,~J_\phi=\frac{1}{2\pi}\oint d\phi~p_\phi.
\end{aligned}
\end{equation}
These expressions give the action variables as functions of the constants of motion, $({\cal H}, E,L_z,Q)$, and hence they are conserved, as indicated in Eq.~\eqref{eq:dotJalpha}. We then obtain the corresponding angle variables using the canonical transformation from $(x^\mu,p_\mu)$ to $(q^\alpha,J_\alpha)$~\cite{Schmidt:2002qk,Hinderer:2008dm}
\begin{align}
p_\mu= \frac{\partial{\cal W}}{\partial x^\mu},~q^\alpha = \frac{\partial {\cal W}}{\partial J_\alpha},\label{eq:canonicaltransform}
\end{align}
where ${\cal W}={\cal W}(x^\mu,J_\alpha)$ is referred to as the Hamilton's characteristic function and is given by~\cite{Carter:1968rr}
\begin{align}
    {\cal W}=- Et+L_Z\phi+\int dr\frac{\sqrt{V_r}}{\Delta}+\int d\theta \sqrt{V_\theta}.\label{eq:CharacteristicW}
\end{align}
Note that the first equation in Eq.~\eqref{eq:canonicaltransform} is already satisfied by virtue of ${\cal W}$ in Eq.~\eqref{eq:CharacteristicW}. The second equation of Eq.~\eqref{eq:canonicaltransform} shows that, once ${\cal W}$ is known, the angle variables~$q^\alpha$ -- and hence $\psi$ in Eq.~\eqref{eq:psi} -- can be evaluated along an orbit. For bound geodesic motion in the unperturbed Kerr spacetime, analytic prescriptions for evaluating $q^\alpha$ are well established \cite{Schmidt:2002qk,Hinderer:2008dm} (see also Appendix~D of the published version of Ref. \cite{Pan:2023wau}).

An action-angle description is generically unavailable for the tidally-deformed Kerr system. Closely following the approximation scheme employed in Ref.~\cite{Pan:2023wau}, we therefore map each numerically obtained phase-space point~$(x^\mu,p_\mu)$ of a tidally-perturbed orbit to the action-angle variables~$(q^\alpha,J_\alpha)$ for the unperturbed Kerr system. Specifically, we first evaluate Eq.~\eqref{eq:CharacteristicW} using the numerically obtained physical coordinates~$(x^\mu,p_\mu)$ of the tidally-perturbed orbit, even though ${\cal W}$ is defined for the unperturbed Kerr system. We then obtain the angles~$q^\alpha$ from the second equation of Eq.~\eqref{eq:canonicaltransform} and hence construct $\psi$. For sufficiently small~$\epsilon$, the resulting angle is expected to serve as a useful proxy for understanding the resonant dynamics of the tidally-deformed orbit.

The left panel of Fig.~\ref{fig:ResonantAngles} shows the evolution of $\psi$ over many orbital cycles for $(n_{\rm res},k_{\rm res}, m_{\rm res})=(3,-2,0)$ in the non-rotating case. The black curve shows $\psi$ along a representative orbit evolved from an initial radius within the plateau  shown in Fig.~\ref{fig:orbit_tide_res_Schwarz}~(and the resonant island shown in Fig.~\ref{fig:PM_1}), whereas the dashed blue and solid red curves correspond to representative results for orbits evolved from initial radii exterior to the radial interval below and above the initial-radius range spanned by the plateaus. For the former orbit, the corresponding~$\psi$ librates with a nearly constant period 
in the range between $-\pi$ and $\pi$, i.e., it is phase-locked. On the other hand, for the latter orbit, the corresponding~$\psi$ circulates monotonically through the full range of $\psi ({\rm mod}~2 \pi N)$, where $N$ is an illustrative integer constant. Thus, the orbits responsible for tidal resonances in non-rotating tidally-deformed EMRIs are those that begin from the interior of the plateau (or equivalently evolve inside the resonant island).

Equation~\eqref{eq:psidot} indicates that for non-resonant orbits, the sign of $d\psi/d\tau$ is determined by the zeroth-order frequency combination that does not change its sign over many orbital cycles. For $m_{\rm res}=0$, the ratio~$\omega_r^{\rm Kerr}/\omega_\theta^{\rm Kerr}$ is smaller than $|k_{\rm res}/n_{\rm res}|$ for orbits evolved below the resonant region associated with the corresponding plateau. Conversely, for orbits above the resonant region, the ratio $\omega_r^{\rm Kerr}/\omega_\theta^{\rm Kerr}$ is larger than $|k_{\rm res}/n_{\rm res}|$, since it increases monotonically with the initial radius. Therefore, the zeroth-order frequency combination, and hence $d\psi/d\tau$, changes sign across the resonant region from negative to positive.

The right panel of Fig.~\ref{fig:ResonantAngles} shows the evolution of $\psi$ in the rotating case for orbits with initial radii within the plateaus shown in Fig.~\ref{fig:orbit_tide_res_Kerr}. We found that $\psi$ circulates along orbits with initial radii outside the plateaus, as performed for the red and dashed blue lines shown in the left panel of Fig.~\ref{fig:ResonantAngles}. The solid gray curve shows $\psi$ with $(n_{\rm res}, k_{\rm res},m_{\rm res})=(3,-2,0)$ along a representative orbit evolved from an initial radius within the first column of plateaus (see caption of Fig. \ref{fig:orbit_tide_res_Kerr}). The dashed gray and solid dark red curves show $\psi$ with $(3,0,-2)$ and $(7,-5,0)$ for the same representative orbit with an initial radius within the second column of plateaus. The resonant angles with $(3,-2,0)$ and $(3,0,-2)$ remain bounded between $-\pi$ and $\pi$ (lower and upper black dashed horizontal lines) and are thus phase-locked. These resonant modes correspond to the dominant resonance, and in particular, are identified by the $2/3$ plateaus in $
\omega_r/\omega_\theta$ and $\omega_r/\omega_\phi$, respectively. In contrast, the evolution of $\psi$ with $(7,-5,0)$, and initial conditions inside the $5/7$ plateaus in $\omega_r/\omega_\theta$ (and its multiples), shown in dark red in the right panel of Fig. \ref{fig:ResonantAngles}, circulates indefinitely with respect to proper time. Thus, the mode $(7,-5,0)$, even though evolved from the interior of the plateau, is a ``pseudo-resonance'', with respect to the dominant tidal resonances $(3,-2,0)$ and $(3,0,-2)$, and does not lead to secular changes in the constants of motion. This behavior is consistent with the suppression of resonances for which $k+m$ is odd, as discussed in Refs.~\cite{Gupta:2021cno,Gupta:2022fbe}.

Another important observation in the rotating case is that the two libration modes have different amplitudes and periods. In the effective pendulum analogy, a small libration amplitude of $\psi$ corresponds to motion near the center of the resonant island, whereas a larger amplitude indicates motion closer to the separatrix \cite{Bronicki:2022eqa}. The difference in periods appears to be qualitatively compatible with the corresponding difference in plateau widths: the period of $\psi$ for $(3,-2,0)$, associated with the $2/3$ plateau in $\omega_r/\omega_\theta$, is longer than that for $(3,0,-2)$, associated with the $2/3$ plateau in $\omega_r/\omega_\phi$.

\subsection{Implication for EMRIs}

The results presented in the previous sections demonstrate a direct connection between plateaus in the rotation curves and the time evolution of the angle variables for the corresponding resonant orbits. These two signatures illuminate complementary aspects of the same tidal resonant dynamics: the former identifies the parameter space associated with resonances, while the latter reveals which resonant mode is responsible for secular changes in the constants of motion. The evolution of the resonant angle and its connection to tidal resonances have been studied in previous work~\cite{Bonga:2019ycj,Gupta:2021cno,Gupta:2022fbe}. By contrast, plateaus in rotation curves are the main focus of the present work that provides a novel phase-space characterization of tidally-enhanced resonances, that has been overlooked in the literature. In particular, together with the action-angle analysis, we identify that the $2/3$ plateaus in $\omega_r/\omega_\theta$ and $\omega_r/\omega_\phi$ in rotation curves correspond to the resonant parameter region responsible for tidal resonances.

\begin{figure*}[t]
\includegraphics[width=\textwidth]{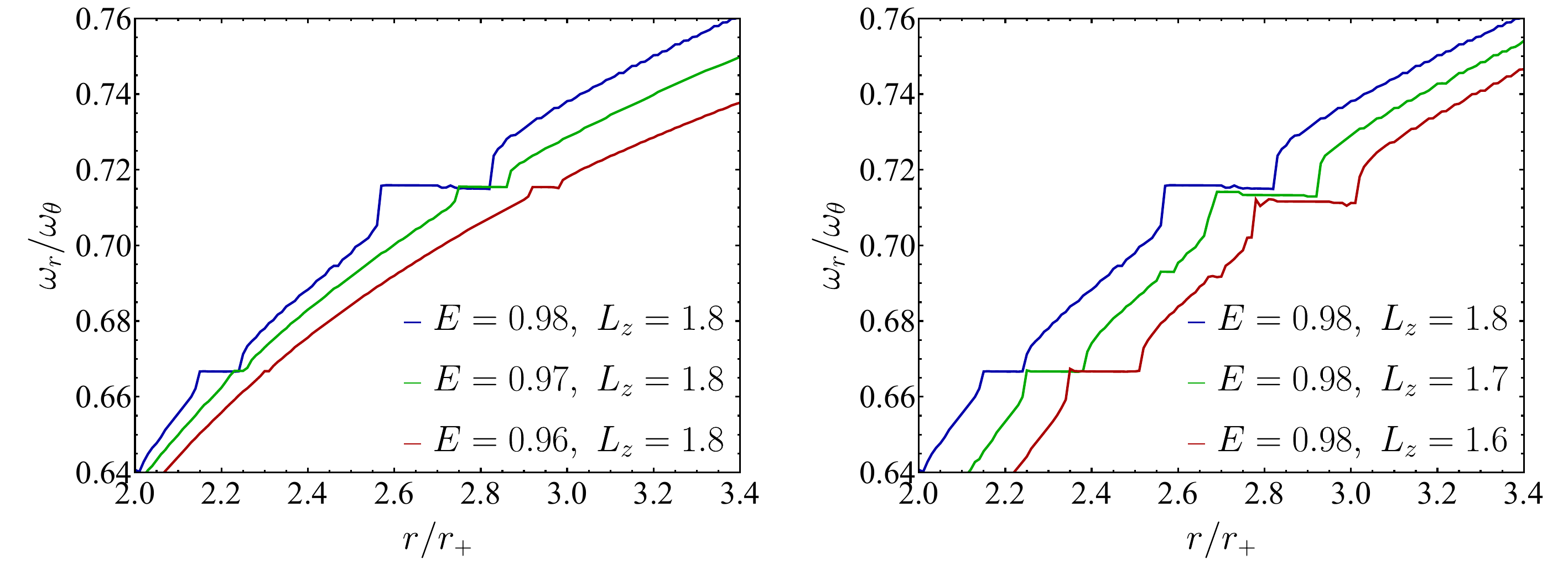}
\caption{The $E/\mu,\,L_z/\mu$ dependence of the rotation curve with a fixed $\epsilon=10^{-8}$, $\phi_{\rm tide}=0$ and $\mu=r_+=1$. The plateaus appear for the $2/3$ and $5/7$ resonances in $\omega_r/\omega_\theta$. It appears that the plateau width tends to increase as $E/\mu$ increases, while the change on $L_z/\mu$ seems to contribute to a shift of the whole rotation curve, while almost keeping the width of the islands intact.}
\label{fig:varyingEandLz}
\end{figure*}

Our findings suggest the following possible scenario for an EMRI in a tidal environment as follows: At some stage of a secondary captured into a bound orbit with a relatively high eccentricity, its orbit may enter one of the finite-width resonant regions identified by the plateaus at its radial turning point. During such an encounter, the orbital motion may be approximated by a bound resonant geodesic, where a tidal resonance can occur. With gradual changes in the constants of motion induced by the tidal perturbation and radiation reaction, both the orbit and the location of the resonant regions are expected to change. In particular, the inner radial turning point of a secondary typically moves inward, while, as shown in Appendix~\ref{appendix:variousparameterset}, decreasing $E/\mu$ and/or $L_z/\mu$ shifts the plateaus outward~(see Fig.~\ref{fig:varyingEandLz}). This indicates that, as the radial interval of the bound orbit and the resonant regions change during an eccentric inspiral, the orbit may intersect finite-width tidal resonant regions associated with different rational frequency ratios more than once. Moreover, plateaus with smaller rational frequency ratios are located closer to the BH horizon. Together with the results from the action-angle description, as an eccentric inspiral proceeds inward, the orbit can encounter tidal resonances at $(n,k,m)=(3,0,-2)$ and then $(3,-2,0)$. 

\section{Conclusions \& Discussion}

Observing a three-body system \cite{Musielak:2014RPPh...77f5901M,Mazzolari:2026A&A}, defined as three celestial bodies of (in-)comparable mass influencing each other's gravitational field, is extremely rare in a highly chaotic form, but very common in general, as well as in hierarchical or restricted forms. While the three-body problem refers to the difficulty of predicting chaotic trajectories, the Universe contains many examples of three or more bodies that behave with enough predictability, even if they technically constitute a three-body, potentially chaotic system. The vast majority of stable, long-lasting three-body systems observed are ``hierarchical'' \cite{Busetti:2018A&A,Lim:2020cvm}, meaning that two celestial bodies are close to each other, while the third one orbits them from a larger distance. These systems act like a two-body system and can remain stable for billions of years. Some examples are triple star systems \cite{Jayla:2018AAS...23143925J}, the familiar Sun-Earth-Moon system, which is a stable three-body system because the Earth-Moon interaction dominates, reducing the Sun's role to a perturbation \cite{Gutzwiller:1998RvMP...70..589G}, and the very recently discovered active triple system of supermassive BHs in the early-Universe galaxy J0148-4214 \cite{Mazzolari:2026A&A}.

In this work, we have examined two restricted three-body problems, i.e., one with an EMRI primary that is non-rotating and another where the supermassive primary is rotating. These results broaden to a full parameter-space analysis those that appeared recently in Ref.~\cite{Katagiri:2026gkz}. We have found that a weak tidal field, even of order $10^{-10}$, can influence the dynamics of the relativistic, restricted, three-body problem significantly, to the point that integrability of geodesics break and the system displays some quite large parts of the parameter space, where signatures of chaos emanate indirectly. 

Even without taking into account the spin of the primary, which usually breaks integrability in non-Kerr EMRIs or Kerr EMRIs within astrophysical environments, we still found imprints of non-integrability through the existence of resonant islands that form and surround stable periodic orbits of the dynamical system, where the periodicity of the central periodic stable point is shared throughout the island. This leads to a clear-cut plateau when one observes the rotation curve of successive orbits, till the separatrix, in particular, resonant ratios of fundamental oscillation frequencies. Even if the observation of fundamental frequencies of a test particle orbiting a tidally-deformed non-rotating BH is practically impossible, it has been shown that the crossing of a resonant island in a variety of non-integrable EMRIs has an explicit effect on the GW emission, i.e. a GW frequency modulation or GW ``glitch'' \cite{Destounis:2021mqv,Destounis:2021rko,Lukes-Gerakopoulos:2021ybx,Deich:2022vna,Mukherjee:2022dju,Polcar:2022bwv,Destounis:2023gpw,Destounis:2023khj,Destounis:2025tjn,Strateny:2025zkd,Azreg-Ainou:2026xcc}, and therefore an overall dephasing and eventual erroneous parameter inference. 

We find that in tidally-perturbed EMRIs with non-spinning primary, 
the resonant island width of the dominant $2/3$ resonance follows a logarithmic relation as the amplitude of the tide increases, in a similar way as it has been observed in Refs.~\cite{Zelenka:2019nyp,Mukherjee:2022dju}, with a slope of $\sim1/2$. When the tidally-deformed EMRI primary is rapidly-rotating then the results become more intricate. Even though there are a plethora of resonances involved in these systems, when the spin of the primary is enabled, we find a seemingly important $5/7$ island, besides the typical $2/3$ ones. In our setup, the plateaus at larger initial radii, associated with the $5/7$ plateau, are wider than those closer to the deformed BH, associated with the $2/3$ plateau.

Another important aspect of our analysis concerns the dependence of the indirect chaotic signatures on the fixed azimuthal angle $\phi_{\rm tide}$ of the adiabatically-moving tertiary. In practice, we ``freeze'' the angle $\phi_{\rm tide}$ to either $0, \pi/2, \pi,$ or $3\pi/2$. We first find an expected symmetry between the results for $\phi_{\rm tide}=0$, $\pi$, and $\phi_{\rm tide}=\pi/2$, $3\pi/2$, due to the $\pi$-periodicity of the quadrupolar tidal perturbation. Setting $\phi_{\rm tide}$ to $0$ or $\pi$ leads to a clear plateau due to the crossing of the $2/3$-resonant island through the radial momentum line $\dot{r}[0]=0$, which is the axis of symmetry of the corresponding Poincar\'e map. On the other hand, when we displace the tidal field at $\phi_{\rm tide}=\pi/2$ or $3\pi/2$, we find that the $2/3$-resonant island becomes lopsided, thus an eccentric test-particle orbit will not cross it if the initial radial momentum is null. A non-zero radial momentum (in the case we studied, $\dot{r}[0]=0.1$) will eventually lead to the crossing of the resonant island and a plateau forms. Thus, the GW glitch behavior should be considered universal in these dynamical systems, no matter the angular initialization of the tidal field.
 
We illustrate that for a fixed tidal field, the increment of the secondary's energy per unit mass, $E/\mu$, leads to wider plateaus in the rotation curve, while the increment of the azimuthal angular momentum per unit mass, $L_z/\mu$, leads to slightly narrower plateaus, together with a visible shift of the rotation curves to the left, i.e. closer to the separatrix~(see Appendix \ref{appendix:variousparameterset}).

The resonant islands observed in the Poincar{\'e} map and plateaus in rotation curves can be interpreted as phase-space manifestation of tidal resonances in EMRIs~\cite{Bonga:2019ycj,Gupta:2021cno,Gupta:2021cno}. Within an action-angle description, we have examined resonant dynamics by tracking the proper-time evolution of the angle variables near the commensurabilities indicated by the plateaus. This establishes whether each resonant angle is phase locked, thereby identifying tidal contribution that can induce secular changes in the constants of motion, the underlying mechanism of tidal resonances. The dominant commensurabilities underlying the $2/3$ plateaus in the radial-to-polar and radial-to-azimuthal frequency ratios identify the parameter region in which the corresponding resonant angle librates over many orbital cycles, hence is phase locked. By contrast, outside these plateaus, the angle combinations associated with these commensurabilities circulate over cycles, indicating that their tidal contributions do not yield secular contributions to the constants of motion. Consistent with Refs.~\cite{Gupta:2021cno, Gupta:2022fbe}, we have also found that an orbit inside the resonant island can still exhibit circulation of the angle combinations associated with a particular commensurability indicated by the $5/7$ plateau in the radial-to-polar frequency ratio. Our results therefore provide a complete non-zero-volume phase-space description of tidal resonances in EMRIs through Poincar{\'e} maps, rotation curves, and the action-angle description.

It is worth noting that Ref.~\cite{Bronicki:2022eqa} proposed a Newtonian-analog model of a Kerr BH with an external tidal field. Our analysis upgrades this Newtonian phase-space study to a relativistic setting formulated directly at the level of the spacetime geometry and related to the properties of the binary companion that sources the external tidal field. By adapting the computational framework of Ref.~\cite{Bronicki:2022eqa} for constructing approximate gravitational waveforms to our model, one could estimate the timescale over which the tidal perturbation affects the orbital dynamics and quantify the region in which tidal effects may need to be included in waveform modeling. Such an extension could also be used to determine which phase-space entry points into a resonance result in substantial tidal effects.

Further studies based on non-Kerr Newtonian analogs \cite{Eleni:2019wav,Eleni:2023mjx}, subject to a generic conservative self force producing the secular motion of orbits, have found that the resonant effect found in studies of relativistic non-Kerr EMRIs, that only include dissipative radiation reaction, can be significantly enhanced when the conservative part of self force is taken into account, thus leading to an overall dynamics where non-integrable tidal resonances become significantly dominant.

The analysis performed here is the first, full-range study of the phase-space of tidally-deformed EMRIs and how/why they break the integrability of geodesics, leading to indirect ``observables'' of chaos in their orbital evolution and subsequently to the gravitational waveforms of such EMRIs, similar to those found in \cite{Destounis:2021mqv}. Even though we have kept our analysis at the conservative level of test-particle geodesics, a possible extension to actual inspirals with appropriate fluxes should reveal exactly how these GW glitches emanate and influence the overall dephasing with vacuum, two-body EMRIs. At the moment, such complicated dynamical systems can only be treated as actual inspirals that emit GWs, through PN techniques for Kerr EMRIs, that have been significantly extended in various works. The actual, Teukolsky-based treatment of tidally-deformed EMRIs is still extremely difficult, especially for generic orbits in three-body dynamics. We leave this lengthy analysis for a future work.
%
\begin{acknowledgments}
The authors would like to thank Vitor Cardoso for his insightful comments on the current manuscript.
K.D. acknowledges financial support provided by FCT–Fundação para a Ciência e a Tecnologia, I.P., under the Scientific Employment Stimulus – Individual Call – Grant No. 2023.07417.CEECIND/CP2830/CT0008. K.D. would also like to thank the Fundação para a Ciência e Tecnologia (FCT), Portugal, for the financial support to the Center for Astrophysics and Gravitation (CENTRA/IST/ULisboa) through grant No. UID/PRR/00099/2025 and grant No. UID/00099/2025.
T.K. is supported by the MUR FIS2 Advanced Grant ET-NOW (CUP:~B53C25001080001) and by the INFN TEONGRAV initiative.
This project has received funding from the European Union’s Horizon MSCA-2022 research and innovation programme “Einstein Waves” under grant agreement No. 101131233.
S. M. is thankful to the Inspire Faculty Grant (DST/INSPIRE/04/2020/001332) from DST, Govt. of India, Prime Minister Early Career Research Grant (ANRF/ECRG/2024/004108/PMS) by ANRF, Govt. of India, and the New Faculty Seed Grant (NFSG/PIL/2023/P3794) provided by BITS Pilani (Pilani), India, for financial support. He (S. M.) is also grateful to the Visiting Associateship Program at IUCAA, Pune, for academic visits where a part of this work was carried out. Finally, S.M. is grateful to the hospitality provided during academic visits at the University of T\"{u}bingen and Sapienza University of Rome, where a part of this work was carried out. 
\end{acknowledgments}

\appendix

\section{Supplemental parameter dependence of rotation curves}
\label{appendix:variousparameterset}

\subsection{Dependence on the energy and azimuthal angular momentum}
 
In addition to the parameters specifying the tidal environment, i.e., the tidal amplitude~$\epsilon$ and the tidal angle~$\phi_{\rm tide}$, our geodesic description is characterized by $E/\mu$ and $L_z/\mu$. First, we discuss how the rotation curves change when $E/\mu$ is varied while fixing $L_z/\mu$. The left panel of Fig.~\ref{fig:varyingEandLz} shows that both the $2/3$ and $5/7$ resonant plateaus in $\omega_r/\omega_\theta$ expand as $E/\mu$ increases. This indicates that increasing $E/\mu$ effectively enhances the importance of the tidal perturbation, in a similar way to increasing the tidal amplitude~$\epsilon$. Another interesting observation is that the rotation curves move rightward as $E/\mu$ decreases. This shift is inherited from the behavior of the rotation curve of unperturbed Kerr geodesics: for fixed $L_z/\mu$, decreasing $E/\mu$ shrinks the radial interval over which bound motion is allowed. 

Next, we discuss the dependence of the rotation curves on $L_z/\mu$. The right panel of Fig.~\ref{fig:varyingEandLz} shows that decreasing $L_z/\mu$ expands the plateaus, in contrast to the behavior when $E/\mu$ decreases. Therefore, within the present setup, the tidal-resonant structures are enhanced by increasing the tidal amplitude~$\epsilon$ and the specific secondary energy~$E/\mu$, as well as decreasing the azimuthal angular momentum~$L_z/\mu$. Similar dependence of chaotic behavior on $E/\mu$ and $L_z/\mu$ has been reported for naked singularities~\cite{Lukes-Gerakopoulos:2010ipp} and slowly rotating compact objects \cite{Destounis:2023cim}. Moreover, as in the case where only $E/\mu$ is varied, decreasing $L_z/\mu$ shifts the rotation curves rightward. This behavior is also inherited from unperturbed Kerr geodesics: for fixed $E/\mu$, a lower azimuthal angular momentum shrinks the radial interval over which bound motion is allowed.

\subsection{Dependence on the primary BH spin}

Finally, we compare the chaotic imprints of tidally-perturbed massive compact objects with respect to the spin of the primary rotating object. We have found the rotation curve widths for varying spin of the primary, since its mass can be scaled out, and observe an enlargement of the islands of stability as the spin $a/M$ increases. More specifically, the most obvious observation is that from non-rotating $a=0$ to rotating tidally-deformed binary with $a/M=0.8$, the width of the $2/3$-island decreases extremely from $0.36r_+$ ($a=0$) to $0.09r_+$ ($a/M=0.8$), that is a $120\%$ of absolute percentage difference. The aforementioned information regarding the $2/3$ plateau widths can be extracted from Fig. \ref{fig:Schwarz_width_vs_eps} ($a=0$) and Fig. \ref{fig:widthvseps} ($a/M=0.8$), where in both cases $\epsilon=10^{-8},\,\phi_{\rm tide}=0,\,E/\mu=0.98$ and $L_z/\mu=1.8r_+$. Even though there exists a seemingly more dominant, i.e., wider, $5/7$-resonant island, that is absent in non-rotating cases, we have shown in the main text that this resonance is not bounded as the $2/3$ one, thus under radiation reaction it will shrink significantly. In any case, the underlying reason why the percentage difference for the $2/3$-island from $a=0$, where $2/3$ is the dominant resonance, to $a/M=0.8$, where $2/3$ seems subdominant, is so vast, and the increment of spin does not seem to enlarge the island width. Nevertheless, the appearance of a new adiabatically-decaying island is possibly the balancing factor between non-rotating and rotating tidally-deformed BHs, in what regards spin effects and corresponding rotation curve plateaus, at the geodesic level.

\bibliography{refs}

\end{document}